\documentclass[a4paper,11pt]{article}
\usepackage{amsmath}
\usepackage{amsfonts}
\usepackage{amssymb}
\usepackage[left=2cm, right=2cm, top=2.5cm, bottom=2.5cm]{geometry}
\usepackage{url}
\usepackage{hyperref}
\usepackage{graphicx}
\usepackage{subcaption}
\usepackage[compat=1.1.0]{tikz-feynman}
\usepackage{array}
\usepackage{cancel}
\usepackage{IEEEtrantools}
 \usepackage{multicol}
 \usepackage{pdfpages}
\usepackage{autobreak}
\usepackage{simpler-wick}
\usepackage{floatrow}

\newcommand{\iu}{{i\mkern1mu}}
\newcommand{\qp}{{\left( q + \delta + \epsilon \right) }}
\newcommand{\de}{{\delta}}
\newcommand{\e}{{\epsilon}}

\graphicspath{{Images/}}

\title{\boldmath Four-point functions in large $N$ Chern-Simons fermionic theories}

\author{Rohit R. Kalloor}
\date{%
    Department of Particle Physics and Astrophysics \\ Weizmann Institute of Science\\Rehovot 7610001, Israel\\ \ \\
    \texttt{e-mail:} \href{mailto:rohit.reghupathy@weizmann.co.il}{\texttt{rohit.reghupathy@weizmann.ac.il}} \\ \ \\
    \today
}

\begin{document} 
\maketitle

\abstract{We compute all four-point functions involving the operators $J_0$ and $J_1$ in large-$N$ Chern-Simons fermionic theories, in the regime where all external momenta lie along the $z$-axis. We find that our result for $\langle J_0 J_0 J_0 J_0 \rangle$ agrees with previous computations, and that the other correlators fall in line with expectations from bootstrap arguments.}

\newpage

% Introduction------------------------------------------------------------------------

\section{Introduction}

\begin{figure}[t]
    \begin{subfigure}[b]{0.45\textwidth}
        \fbox{ \includegraphics[width=\textwidth]{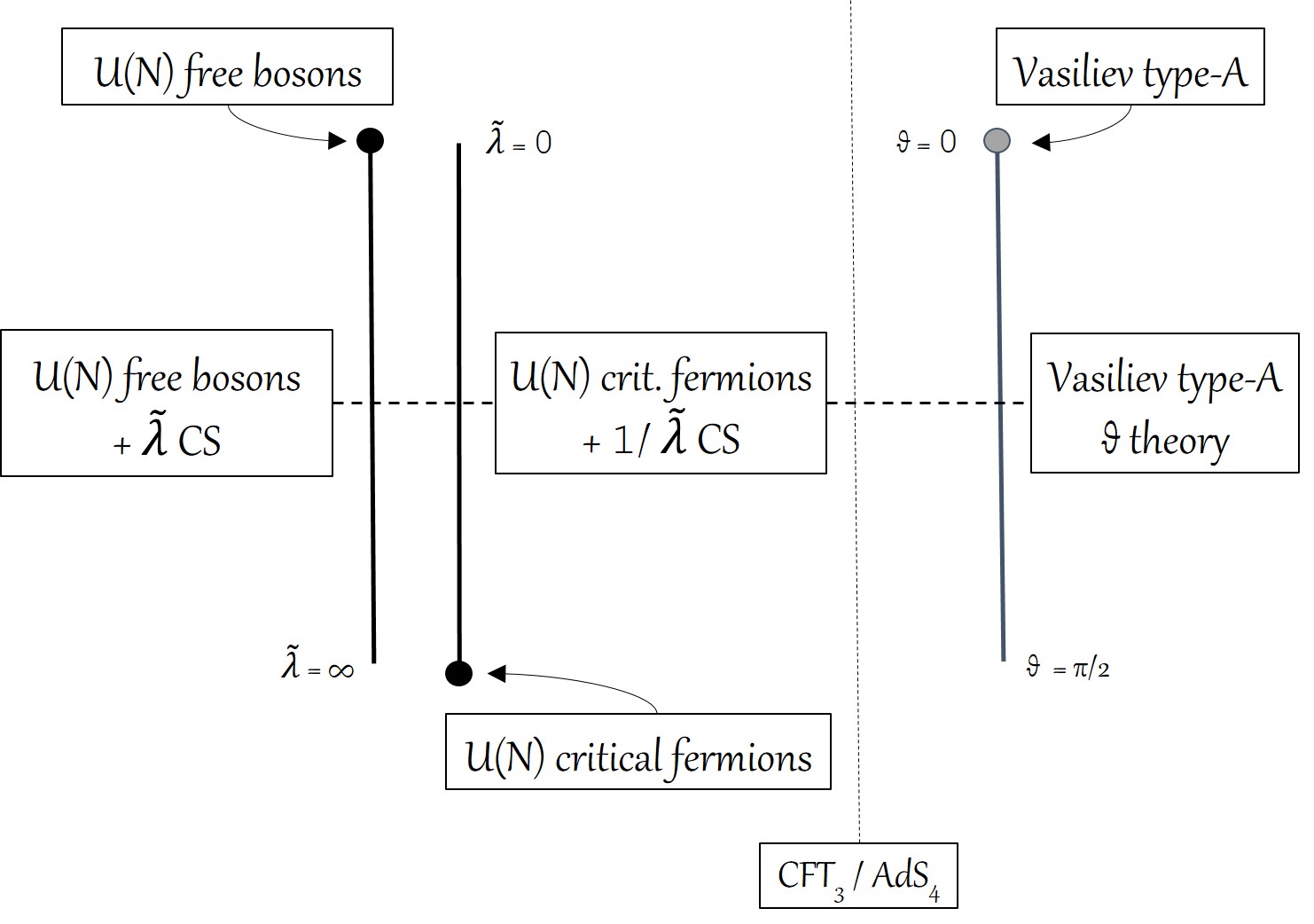} }
        \caption{The quasi-bosonic theories}
        \label{fig:qb}
    \end{subfigure}
    \quad
    ~ %add desired spacing between images, e. g. ~, \quad, \qquad, \hfill etc. 
    \begin{subfigure}[b]{0.454\textwidth}
        \fbox{ \includegraphics[width=\textwidth]{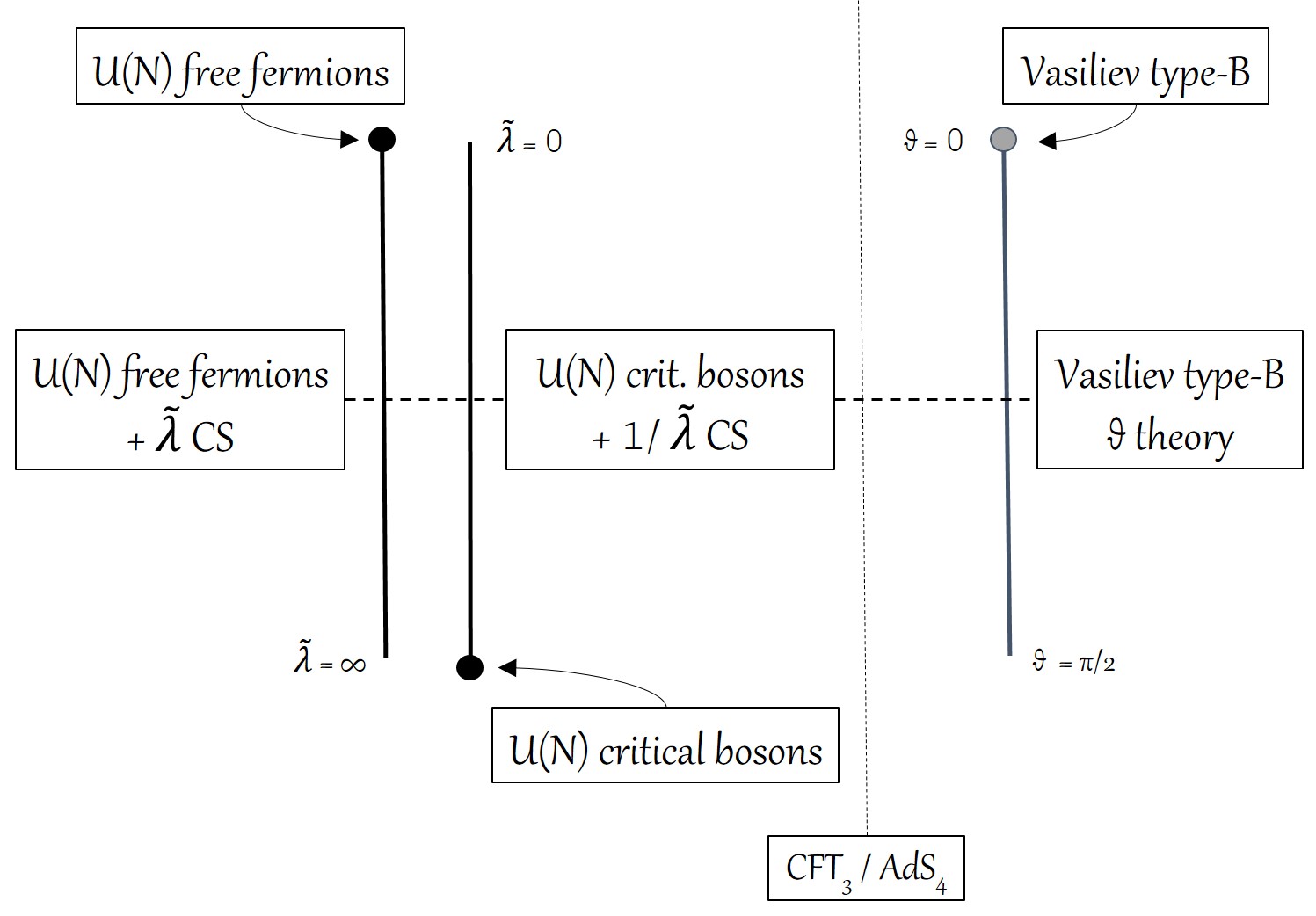} }
        \caption{The quasi-fermionic theories}
        \label{fig:qf}
    \end{subfigure}
    \caption{The bosonisation dualities have two branches that relate two pairs of CS-matter theories: \newline
     (a) \textbf{The quasi-bosonic theories}: the starting point is the singlet sector of the free $U(N)$ bosonic theory (`$U(N)$ free bosons'). We can then turn on the coupling $\tilde{\lambda}$ (a monotonic function of the 't Hooft coupling $\lambda $; see Section \ref{sec:res}), and on taking it to infinity (the strong coupling limit $\lambda \rightarrow 1$), the correlators of the CS-bosonic theories go over to that of the critical fermionic (Gross-Neveu) fixed point, which is the end-point of the line. The bosonisation duality relates this (critical) fermionic theory coupled to CS theory with strength 1/$\tilde{\lambda}$ to the free bosonic theory coupled to CS theory with strength $\tilde{\lambda}$. The corresponding bulk duals are the type-A $\theta$ theories (parametrised by $\theta \in \left[ 0, \ \pi / 2 \right]$). \newline
     (b) \textbf{The quasi-fermionic theories}: starting from the free fermionic theory, we can turn on the CS  coupling as in the previous case, and sending it to infinity gives the critical bosonic (Wilson-Fischer) theory. The relevant bosonisation duality relates free fermionic CS theories to critical bosonic theories coupled to CS. The bulk duals are given by the type-B $\theta$ theories. In this paper, we will be interested in this family of theories.}\label{fig:bosonisation} 
\end{figure}

Three dimensional gauge theories exhibit a rich class of behaviours, thanks in no small part to Chern-Simons (CS) interactions. The topological degrees of freedom brought in by the CS sectors, originally used to create massive gauge bosons, also give rise to features such as anyons and non-trivial ground state manifolds (see \cite{Dunne:1998qy,Witten:2015aoa, Tong}). As was recently found, there is a large network of dualities -- IR and otherwise, that relate such theories\cite{Seiberg:2016gmd,Karch:2016sxi}. One such class of dualities are the 3d bosonisation dualities\cite{Aharony:2012nh,Aharony:2011jz,Giombi:2011kc, Aharony:2015mjs} (see Fig. \ref{fig:bosonisation}), which are analogous to the bosonisation duality in two dimensions. Roughly speaking, these dualities connect bosons to fermions via particles that have fractional statistics, and hence "interpolate" between the two.

The original motivation for the bosonisation dualities came from looking at the bulk gravity duals of these theories, which are the higher-spin gravity theories of Vasiliev on $AdS_4$ (see \cite{Giombi:2016ejx} for a review and references). These theories, being constrained by the powerful, infinite-dimensional higher-spin symmetry, are believed to be the classical limits of one of the more tractable examples of quantum gravity. This symmetry is realised in the boundary CFT via a set of conserved, higher-spin currents. But while its precise effects in the bulk remain an open puzzle,  Maldacena and Zhiboedov \cite{Maldacena:2011jn} showed that it has a trivialising effect on the boundary CFT\footnote{Since the most rigorous definition we have of a theory of quantum gravity is furnished by the $AdS/CFT$ correspondence, this means that in a sense, \cite{Maldacena:2011jn} have solved higher-spin gravity. However, it is still a formidable challenge to realise this solution in terms of the variables of the bulk theory.}, i.e., any CFT satisfying certain assumptions, and with a set of conserved, higher-spin currents, is free. Thus, higher-spin symmetry in its full glory, is uninteresting from a field theory point of view since it's merely a symmetry of the free theory.

However, for CFT's with a large-$N$ expansion, one may allow these conservation laws to be flouted at $\mathcal{O}(1/N)$ (weakly broken higher-spin symmetry)\footnote{In the bulk, this corresponds to a change in the boundary conditions, breaking the corresponding (gauge) symmetry in the bulk. The higher-spin fields get a mass at $\mathcal{O}(1/N)$ and the corresponding boundary currents are not conserved starting at $\mathcal{O}(1/N)$. Hence the name `weakly/slightly broken higher-spin symmetry'.};  this opens up a bit more room for interesting dynamics, while still leaving in place powerful constraints. For instance, as shown yet again by Maldacena and Zhiboedov \cite{Maldacena:2012sf}, the three-point functions of all single-trace operators are fixed to $ \mathcal{O} (1/N )$, up to one or two parameters. Examples for theories of this type are the CS-matter theories.

At this point, a natural question to ask would be if weakly broken higher-spin symmetry is powerful enough to constrain all the correlators of the theory (at least to leading order in $1/N$); this would mean that CS-matter theories are the only examples of theories with such a symmetry. Answering this question directly would involve working with the corresponding anomalous higher-spin Ward identities, which are complicated integro-differential equations that relate the various correlators; and is a formidable task. Possible avenues of attack would be to express these relations in momentum space, where they are algebraic (as was noted by \cite{Turiaci:2018dht}); or stick to position space, and move to the conformal frame, where certain simplifications occur. The second approach was used successfully by \cite{Li:2019twz} to fix certain four-point functions, and hence provide some evidence toward answering this question in the affirmative. Yet another way of doing these computations was explored in \cite{Skvortsov:2018uru}.  

%\footnote{$J_0$ and $J_2$ are, respectively, the scalar and the stress-energy tensor in these theories and will be defined in Section \ref{}.}

At any rate, the mere presence of such powerful constraints means that we may expect a lot of simplifications while doing computations in CS-matter theories. Thus, there may be some hope of constructing a complete solution to these theories to leading order\footnote{At this order, one may also take advantage of the simplifications that come with large-$N$.} in $1 / N $. Indeed, it was found (\cite{Aharony:2012nh,Aharony:2011jz,Giombi:2011kc}) that at leading order in $ 1/ N$, correlation functions may be systematically computed to all orders in perturbation theory in the 't Hooft coupling $ \lambda = N / k $, in momentum space, in the regime where the external momenta are pointed along the z-axis (in the following, we shall refer to this formalism as ``collinear perturbation theory"; see Section \ref{sec:4pt}). This technology was used to compute two-, three- and even four-point functions\footnote{\cite{Bedhotiya:2015uga} were the first to use this formalism to compute the four-point function $\langle J_0 J_0 J_0 J_0 \rangle$, but their result was incorrect and was rectified by \cite{Turiaci:2018dht}.} of single-trace operators in these theories and the results were found to depend on the 't Hooft coupling in a simple manner. There was also  the work of \cite{Yacoby:2018yvy}, where a recipe to compute all $n-$point functions of the form $ \langle J_0 J_0 ... \rangle $ in the collinear regime (to leading order in $1/N$), was given.

However, beyond the three-point level, this approach loses much of its power for a simple reason. Up to the three-point level, conformal symmetry on its own fixes all correlators up to constant factors (the OPE coefficients). Then, provided the respective tensor structures don't vanish in the collinear limit (as was the case for \cite{Aharony:2012nh,GurAri:2012is}), the collinear calculation determines the full result. This is no longer the case at the four-point level, and thus the first full calculation of a four-point function $\langle J_0 J_0 J_0 J_0 \rangle$ by \cite{Turiaci:2018dht} relied on a more powerful technique -- the OPE inversion formula \cite{Caron-Huot:2017vep}. The inversion formula was first used to bootstrap the correlator and fix it up to three undetermined constants:
\begin{IEEEeqnarray}{rCl}
\langle J_0 J_0 J_0 J_0 \rangle &^{\text{quasi-fer}} = & \frac{1}{ \tilde{N}} \left( \frac{1}{ ( 1 + \tilde{ \lambda }^2 )^2 } \langle J_0 J_0 J_0 J_0 \rangle _{ \text{fer}} + b_1 G^{AdS}_{\phi^4} +  b_2 G^{AdS}_{ \left( \partial\phi \right)^4} + b_3 G^{AdS}_{\phi^2 \left( \partial ^3 \phi \right)^2} \right)
\label{eqn: scalar_bootstrap}
\end{IEEEeqnarray}
where $\tilde{\lambda}$ parametrises the quasi-fermionic theories and is monotonically related to the `t Hooft coupling (see Section \ref{sec:res} for details); and the $G^{AdS}$'s are 4-pt Witten diagrams in $AdS$; a numerical evaluation of the four-point function (using collinear perturbation theory) for various values of the external momenta then gave linear equations for the $b_i$'s, which were solved to set these constants to 0.

The second instance of such a calculation was the recent work of \cite{Li:2019twz}, where they calculated the four-point functions $\langle J_0 J_0 J_0 J_0 \rangle $ and $\langle J_2 J_0 J_0 J_0 \rangle $\footnote{$J_0$, $J_1$, and $J_2$ are respectively the scalar, the conserved spin-1 current, and the stress-energy tensor in these theories and will be defined in Section \ref{sec:4pt}.} purely from the (anomalous) higher-spin Ward identities.

%both cases, the exact results were obtainable due to the full tensor structures being available from soluble theories, such as the free theory and the critical $O(N)$ model. More precisely, in the quasi-fermionic theories,
%\begin{IEEEeqnarray}{rCl}
%\langle J_0 J_0 J_0 J_0 \rangle & = & \frac{1}{ \tilde{N}} \frac{1}{ ( 1 + \tilde{ \lambda }^2 )^2 } \langle J_0 J_0 J_0 J_0 \rangle_{\text{fer}} \\
%\langle J_2 J_0 J_0 J_0 \rangle & = & \frac{1}{ \tilde{N}} \left( \frac{1}{ ( 1 + \tilde{ \lambda }^2 )^2 } \langle J_2 J_0 J_0 J_0 \rangle_{\text{fer}} 
%+
%\frac{\tilde{ \lambda }}{ ( 1 + \tilde{ \lambda }^2 )^2 } \langle J_2 J_0 J_0 J_0 \rangle_{\text{bos}} 
%\right)
%\end{IEEEeqnarray}
% In the case of $\langle J_0 J_0 J_0 J_0 \rangle$, the exact result was proportional to the free theory, while $\langle J_2 J_0 J_0 J_0 \rangle$, there were two structure, one from the free theory and one from the critical $O(N)$ model.
%For the cases discussed above, the $\tilde{ \lambda } $ dependence of the four-point functions are of the form expected from the three-point functions.

The aim of this paper is to probe the $ \tilde{\lambda } $ dependence of two other spinning correlators -- namely $\langle J_1 J_0 J_1 J_0 \rangle $ and $\langle J_1 J_1 J_1 J_1 \rangle $, in $SU(N)_k$ Chern-Simons theory with a fundamental fermion, via collinear perturbation theory. We compute these correlators analytically in the collinear limit and verify that the $\tilde{\lambda}$ dependence is exactly as anticipated from the three-point functions via bootstrap arguments (see Section \ref{sec:res}). We note that this is the first analytic result that displays the (parity-) odd tensor structures that cannot be obtained from various soluble theories\footnote{The tensor structures that appear in $ \langle J_0 J_0 J_0 J_0 \rangle $ and $ \langle J_2 J_0 J_0 J_0 \rangle $ may be computed in the free fermionic / critical bosonic theories, and hence the computations described earlier don't have this feature.}, albeit in a special kinematic regime. However, we will have difficulties applying the methods of \cite{Turiaci:2018dht} to our problem, since we will not be able to construct an ansatz for the full correlator due to these odd tensor structures. That is, we will find by a direct computation that (in the collinear limit):
\begin{align}
\langle \tilde{J}_1 \tilde{J}_0 \tilde{J}_1 \tilde{J}_0 \rangle  & \sim  \frac{1}{\tilde{N}} \left(  \frac{1}{(1 + \tilde{\lambda}^{2} )^2 }  \langle \tilde{J}_1 \tilde{J}_0 \tilde{J}_1 \tilde{J}_0 \rangle _{\text{fer}} +  \frac{\tilde{\lambda}}{( 1 + \tilde{\lambda}^{2} )^2 }  (\#) +  \frac{\tilde{\lambda}^2 }{( 1 + \tilde{\lambda}^{2} )^2 }  \langle \tilde{J}_1 \tilde{J}_0 \tilde{J}_1 \tilde{J}_0 \rangle _{\text{bos}} \right) \\
\langle \tilde{J}_1 \tilde{J}_1 \tilde{J}_1 \tilde{J}_1 \rangle  & \sim  \frac{1}{\tilde{N}} \left(  
\frac{1}{(1 + \tilde{\lambda}^{2} )^2 } \langle \tilde{J}_1 \tilde{J}_1 \tilde{J}_1 \tilde{J}_1 \rangle _{\text{fer}} 
+  \frac{\tilde{\lambda}}{( 1 + \tilde{\lambda}^{2} )^2 }  (\#) 
+  \frac{\tilde{\lambda}^2 }{( 1 + \tilde{\lambda}^{2} )^2 }  (\#) 
+ \frac{\tilde{\lambda}^3 }{( 1 + \tilde{\lambda}^{2} )^2 } (\#) \right.
\nonumber \\
& \ \ \ \ \ \ \ \ \ \ \ \ \ \ \ \ \ \ \ \ \ \ \ \ \ \ \ \ \ \ \ \ \ \ \ \ \ \ \ \ \ \ \ \ \ \ \ \ \ \left. + \frac{\tilde{\lambda}^4 }{( 1 + \tilde{\lambda}^{2} )^2 } \langle \tilde{J}_1 \tilde{J}_1 \tilde{J}_1 \tilde{J}_1 \rangle _{\text{bos}}
\right)
\end{align}
with $ \#$'s standing in for the extra structures that do not come from soluble theories\footnote{There is a small caveat to this, in that our results point to a relation between the `fer', `bos', and the middle ($ \tilde{ \lambda } ^2 $) structure in $ \langle J_1 J_1 J_1 J_1 \rangle $ (see Section \ref{sec:res}). However, we believe that this is an artefact of the collinear regime, and do not know if it holds
 in the general case.}; we know them for collinear momenta, but can't compute these structures in the general case. Hence, we do not know the full analogue the first term in Eqn. (\ref{eqn: scalar_bootstrap}). 
 
But we can say with some confidence that the methods of \cite{Turiaci:2018dht} should be applicable in this case as well. Specifically, in the case of $ \langle J_1 J_1 J_1 J_1 \rangle $, the $AdS$ diagrams that can contribute extra terms as in (\ref{eqn: scalar_bootstrap}) are known\footnote{We thank S. Minwalla for this information, based on \cite{Minwalla:contact}.}, and we have verified that they do not vanish in the collinear limit. Hence, this correlator may be fully determined from the analytic-in-spin answer (the analogue of the first term in Eqn. (\ref{eqn: scalar_bootstrap})) and our result. As for $ \langle J_0 J_0 J_1 J_1 \rangle $, the relevant allowed diagrams are not known, but we have verified that no contact diagram (to second order in derivatives) vanishes in the collinear limit (see Section \ref{sec:res}). We thus conjecture that in both of these cases, as in $\langle J_0 J_0 J_0 J_0 \rangle $, the result is analytic in the spin of the intermediate states, with no extra contact-term-in-$AdS$ contributions.

The arrangement of the rest of the article is as follows: Section \ref{sec:4pt} reviews some facts about CS-matter theories, gives a brief summary of the formalism of collinear perturbation theory, and gives the details of our calculation of the four-point functions; Section \ref{sec:res} lists our results and gives a more complete discussion of the issue of $ \tilde{ \lambda }$-dependence that was mentioned here.

% Section 2--------------------------------------------------------------------------- 

\section{Computing $\left< J_1 J_1 J_0 J_0 \right>$ and other correlators}
\label{sec:4pt}

In this section, we describe the computation of $\langle J_1 J_1 J_{0} J_{0}\rangle$ and other four-point functions to leading order in $1/N$ and all orders in the 't Hooft coupling $\lambda$, for configurations where all the external momenta lie along the z-axis. For concreteness, we shall be talking about a particular correlator $\langle \langle J_1 ^{+}(q + \epsilon + \delta) J_1 ^{-}(-q) J_{0}(-\epsilon) J_{0}(-\delta)\rangle \rangle$ (where $J_1 ^{ \pm } $ are the components of $J_1$, expressed in the light-cone gauge defined below) but the same steps \textit{mutatis mutandis}, give all correlators of the form $\langle J_{1/0} J_{1/0} J_{1/0} J_{1/0}\rangle$. As was noted earlier, $\left< J_0 J_0 J_0 J_0 \right>$ was computed via similar methods in \cite{Turiaci:2018dht}.

\subsection{Generalities}

Chern-Simons (free) fermionic theory\footnote{We use `free fermionic theory' to denote $N$ free fermions coupled to an $SU(N)$ Chern-Simons (gauge) theory. `Free theory' is a theory of free fermions. `Weak', `strong', and `perturbative' describe the 't Hooft coupling. } with an $SU(N)$ gauge group is defined by the (Euclidean) action:
\begin{align}
	\mathcal{L}^{\text{ff}}_{\text{CS}} & =  \iu \frac{k}{4 \pi} \text{Tr} \left( A\ \text{d} A + \frac{2}{3} A^3 \right) + \bar{\psi }\ \cancel{\mathcal{D}}\ \psi \label{eqn:cs_action} \\
	& =  - \iu \frac{N}{8 \pi \lambda} \epsilon^{\mu \nu \rho} \left( A^{\text{a}} _{\mu} \partial_{\nu} A^{\text{a}} _{\rho} + \frac{1}{3} f^{\text{abc}} A^{\text{a}} _{\mu} A^{\text{b}} _{\nu} A^{\text{c}} _{\rho} \right) + \bar{\psi }\ \cancel{\mathcal{D}}\ \psi 
\end{align}
In our case, the fermions $\psi$ are in the fundamental representation of $SU(N)$. As is well known~\cite{Dijkgraaf:1989pz, Deser:1982vy}, this action (\ref{eqn:cs_action}) is gauge-invariant if the coupling $ k $ is an integer, up to a shift of half due to the parity anomaly~\cite{Redlich:1983kn} (which is irrelevant in the large $N$ limit). The spectrum was derived in \cite{Giombi:2011kc}; the primaries are classified via large-N factorisation. The single-trace primaries are denoted by $J_s$, and there is one for each spin $s \in {0, 1, ... }$; these are given (schematically) by:
\begin{align}
J_0 & =  \bar{\psi} \psi \\ 
J_1 ^{\mu} & =  \iu \bar{\psi} \gamma ^{\mu} \psi \\
J_s ^{\mu _1 \mu _2 ... \mu_s }& = \bar{\psi} \gamma ^{\left( \mu _1 \right. } \mathcal{D} ^{\mu _2  } ... \mathcal{D} ^{\left. \mu _s \right) } \psi - \text{(traces)} 
\end{align}
where the parentheses stand for symmetrisation. It was also shown in \cite{Giombi:2011kc} that the dimensions of these operators are independent of $\lambda$ (to leading order in $1/N$ \cite{Giombi:2016zwa, Jain:2019fja}), and in the free fermionic theory, given by:
\begin{equation}
	\Delta _s = \begin{cases} 
      2 + \mathcal{O}\left( \frac{1}{N} \right) & s = 0 \\
      s + 1 + \mathcal{O}\left( \frac{1}{N} \right) & s > 0 
   \end{cases}
\end{equation}
In the large $N$ limit, the double-trace operators are given by a product of two single-trace operators. We shall not have to worry about the other multi-trace operators at this order in $\frac{1}{N}$.

The single-trace primaries for $s = 1, 2$ are the usual conserved current and stress-energy tensor; for $s > 2$, they are the almost conserved higher-spin currents:
\begin{equation}
 \partial . J_s \sim \mathcal{O}\left( \frac{1}{N} \right)
 \label{eqn:cons}
 \end{equation}
where the right-hand side is made up of double- and triple-trace operators (to $\mathcal{O}(\frac{1}{N^2})$, as described in \cite{Giombi:2011kc}) . In the dual gravity theory, this means that the higher-spin fields are massive but the mass is generated at the quantum level (i.e., by loop corrections). As demonstrated by \cite{Maldacena:2012sf}, (\ref{eqn:cons}) leads to great simplifications -- for instance, all the three point functions are fixed to $\mathcal{O}(1/N^2 )$. As discussed in the introduction, it is known that there are at least two families of conformal theories (up to the conjectured dualities) that exhibit weakly broken higher-spin symmetry: the quasi-bosonic and -fermionic CS theories, which are related via dualities to the higher-spin gravity theories of Vasiliev. Whether these are the only examples of such theories; i.e., whether weakly broken higher-spin symmetry, is powerful enough to fix all correlators (to the relevant order in 1/$N$), remains to be seen.

\subsection{Gauge fixing, regulation, computations}

To perform computations, we first move to light-cone coordinates: $ x^{\pm} =  x_{\mp}  =  \frac{1}{\sqrt{2}} \left( x^1 \pm \iu \ x^2 \right) $, and gauge fix the theory by setting $A_- = 0$ as in \cite{Giombi:2011kc} and \cite{Aharony:2012nh}\footnote{This should be viewed as the analytic continuation of the same in the Lorentzian theory, because otherwise, $A_+ = (A_-)^* = 0$.}. This is called the light-cone gauge, and kills the self-interactions between the gluons. The gluon propagator is then given by:
\begin{align}
\left< A_{\mu}^{a} (-p) A_{\nu}^{b} (q) \right> & = (2 \pi)^{3} \delta ^3 (p-q) \delta ^{ab} \  G_{\nu \mu} (p)  \\
G_{+3} (p) & = -G_{3+} (p) = \frac{4 \pi i}{k} \frac{1}{p^+} 
\end{align}
We also need a prescription to regulate the theory; we use the hybrid prescription as in \cite{Aharony:2011jz}, which is a cutoff $\Lambda$ in the $x^1 - x^2 $ plane, and dimensional regularisation in the $x^3$ direction.

The Feynman rules of the theory are:
\begin{figure}[h]
 \includegraphics[scale=0.4]{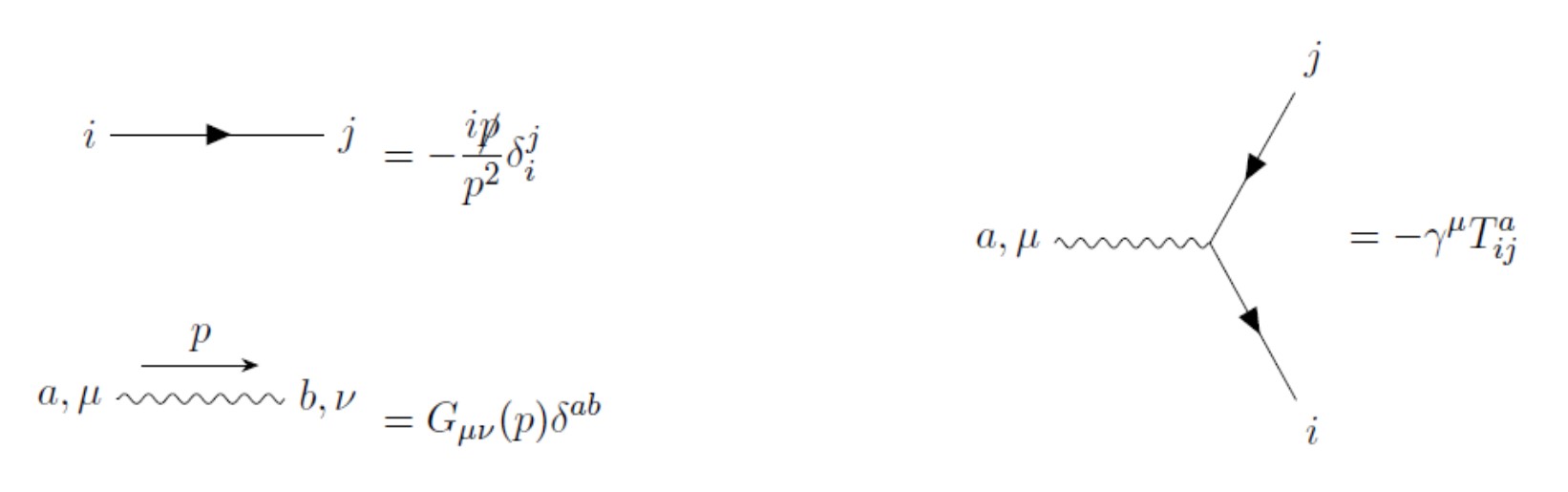} 
\end{figure}

The normalisation of the generators is fixed by $\text{Tr} \left( T^a T^b \right) = -\frac{1}{2} \delta ^{a b} $. We also note that $T^a T^a = C_2 (N) \mathbf{1} = \frac{N}{2} \mathbf{1} + \mathcal{O}\left( \frac{1}{N} \right) $. Using these rules, one may compute most of the ingredients necessary for calculating four-point functions -- namely, the exact $J_0$ and $J_1$ vertices, and the exact fermionic propagator to leading order in $1/N$. This was done in \cite{GurAri:2012is}, and we give the expressions here for completeness. 

The fermionic propagator:
\begin{IEEEeqnarray}{rCl}
\left< \psi_i (p) \bar{\psi} ^j (-q) \right> & = & (2 \pi)^{3} \delta ^3 (p-q)\delta _i ^j \ S(p),  \\ 
S(p) & = & \frac{-i \gamma_{\mu} p^{\mu} +  \lambda p_s \mathbf{1} + i \lambda ^2 p^- \gamma ^+ }{p^2} 
\end{IEEEeqnarray}

The spin-0 vertex:
\begin{IEEEeqnarray}{rCl}
\left< J_0 (-q) \psi _i(k) \bar{\psi}^j (-p) \right> & = & (2 \pi)^{3} \delta ^3 (q+p-k)\delta _i ^j \ V(q,p)  
\end{IEEEeqnarray}
\begin{IEEEeqnarray}{rCl}
V(q,p) & = &  \frac{2 \lambda p^+}{p_s} g \left( \frac{2 p_s}{\left| q \right|} \right) \gamma ^+ + f \left( \frac{2 p_s}{\left| q \right|} \right) \mathbf{1}  \\ 
f(y) & = & \frac{1 + e^{-2 i \widehat{\lambda} \  \text{tan}^{-1}(y)}}{1 + e^{-2 i \widehat{\lambda}\ \text{tan}^{-1}(\Lambda ')}}  \\
g(y) & = & \frac{1 - i \widehat{\lambda} y - (1 + i \widehat{\lambda} y) e^{-2 i \widehat{\lambda} \  \text{tan}^{-1}(y)}}{\widehat{\lambda} y\left( 1 + e^{-2 i \widehat{\lambda} \ \text{tan}^{-1}(\Lambda ')} \right) }  
\end{IEEEeqnarray}

The spin-1 vertex:
\begin{IEEEeqnarray}{rCl}
\left< J_1 ^{\pm} (-q) \psi _i(k) \bar{\psi}^j (-p) \right>  & = & (2 \pi)^{3} \delta ^3 (q+p-k)\delta _i ^j \ V^{\pm}(q,p) 
\end{IEEEeqnarray}
\begin{IEEEeqnarray}{rCl}
V^+(q,p) & = & g^{(+)} \left( \frac{2 p_s}{\left| q \right|} \right) \gamma ^+ + \frac{2p^+}{q} f^{(+)} \left( \frac{2 p_s}{\left| q \right|} \right) \mathbf{1}  \\
f^{(+)}(y) & = & \frac{i}{2}\left( 1 - e^{2 i \widehat{\lambda} \ \left( \text{tan}^{-1} (\Lambda ' ) - \text{tan}^{-1} (y ) \right)} \right)  \\
g^{(+)}(y) & = & \frac{i}{2}\left( 1 - i \widehat{\lambda} y +\left( 1 + i \widehat{\lambda} y \right) e^{2 i \widehat{\lambda} \ \left( \text{tan}^{-1} (\Lambda ' ) - \text{tan}^{-1} (y ) \right)} \right)  \\
V^-(q,p) & = & i \gamma ^- +  \frac{2 \left( p^- \right)^2}{p_s ^2} g^{(-)} \left( \frac{2 p_s}{\left| q \right|} \right)  \gamma ^+ + \frac{2p^-}{q} f^{(-)}\left( \frac{2 p_s}{\left| q \right|} \right) \mathbf{1}   \\
f^{(-)}(y) & = & \frac{i}{2 y^2}\left( 1 - 2 i \widehat{\lambda} y -  e^{-2 i \widehat{\lambda} \ \text{tan}^{-1} (y)} \right)  \\
g^{(-)}(y) & = & - \frac{i}{2y^{2}}\left( 1 - i \widehat{\lambda} y  - \left( 1 + i \widehat{\lambda} y \right) e^{ - 2 i \widehat{\lambda} \ \text{tan}^{-1} (y )} \right)  
\end{IEEEeqnarray}
where we've defined $\Lambda ' = \frac{2 \Lambda}{\left| q \right|}$, and the ``parity-invariant" coupling $\widehat{\lambda} = \text{sgn}(q)\lambda $.

The final piece that we'll need is the four-fermion vertex. This was first computed by \cite{Inbasekar:2015tsa}, but their derivation used conventions different from ours. Since it is important to our calculations, we give our version in Appendix \ref{apd:four_vertex}.

\subsection{The diagrams}

\begin{figure}
    \centering
    \begin{subfigure}[b]{0.24\textwidth}
        \includegraphics[width=\textwidth]{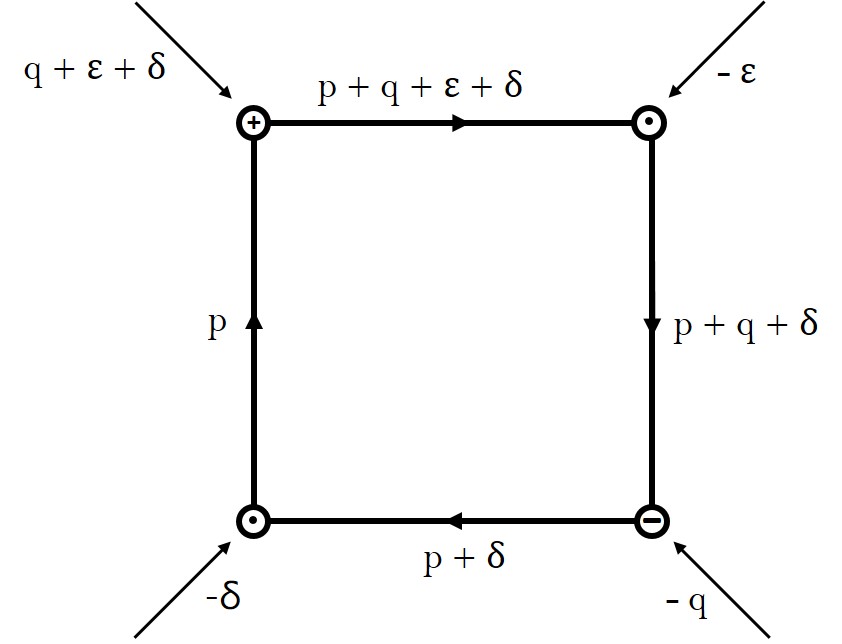}
        \caption{Aa1}
        \label{fig:Aa1}
    \end{subfigure}
    \begin{subfigure}[b]{0.24\textwidth}
        \includegraphics[width=\textwidth]{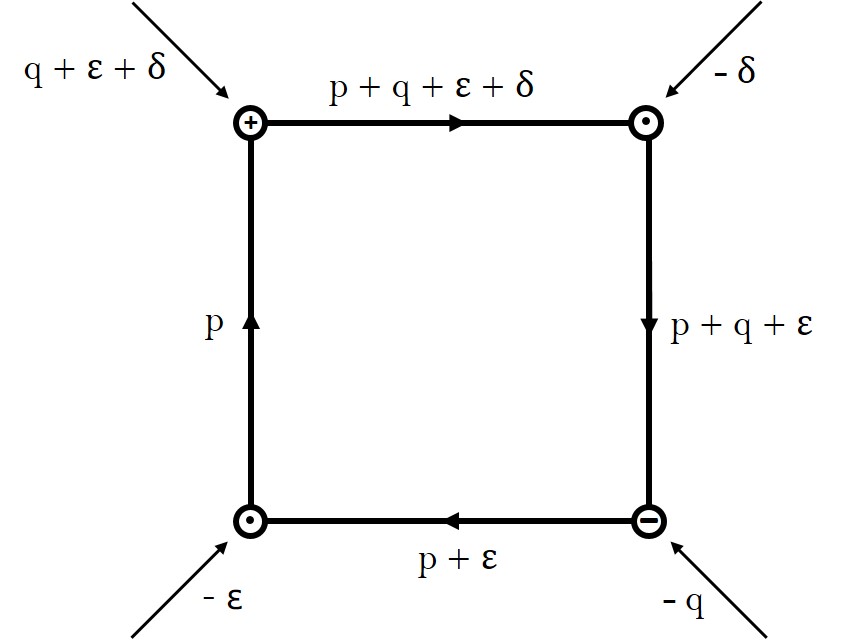}
        \caption{Aa2}
        \label{fig:Aa2}
    \end{subfigure}
    \begin{subfigure}[b]{0.24\textwidth}
        \includegraphics[width=\textwidth]{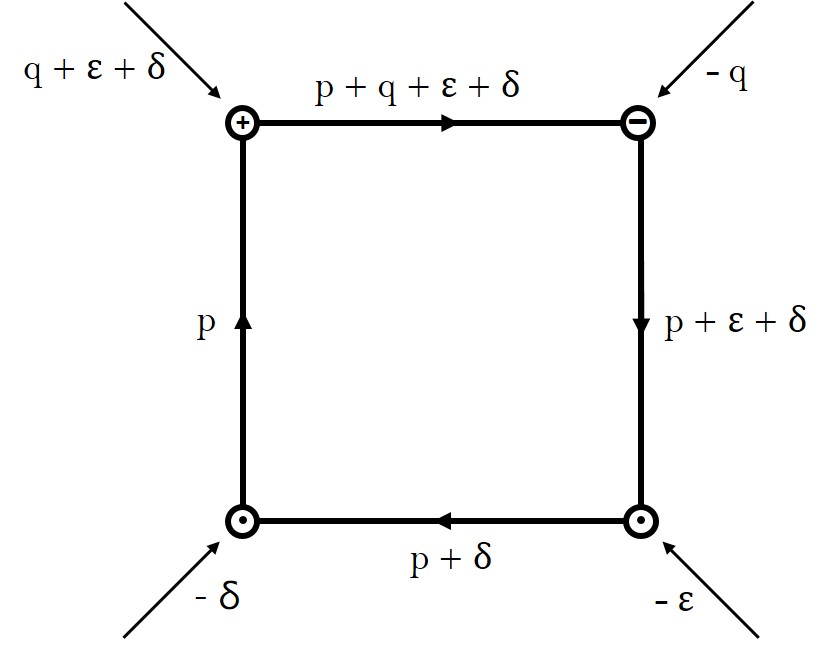}
        \caption{Ab1}
        \label{fig:Ab1}
    \end{subfigure}
    \begin{subfigure}[b]{0.24\textwidth}
        \includegraphics[width=\textwidth]{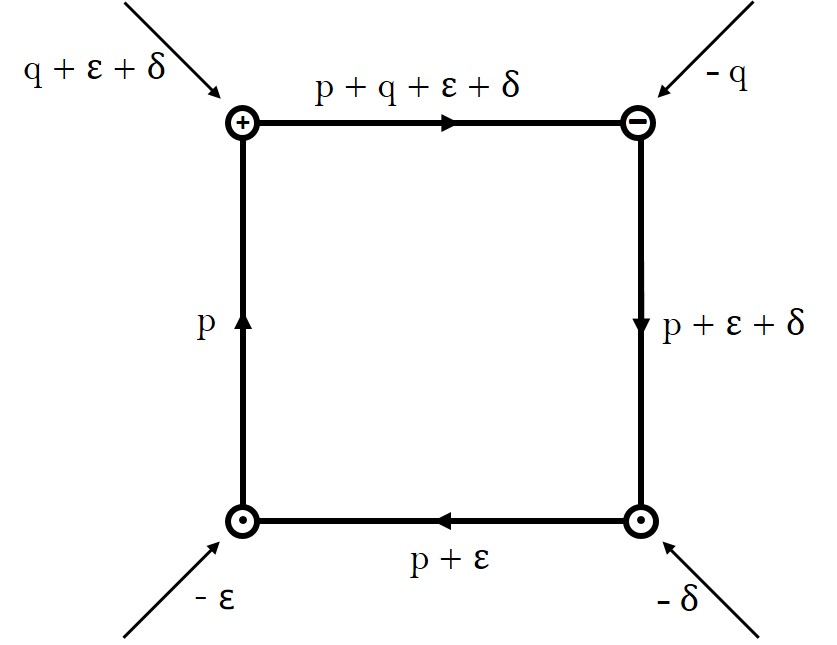}
        \caption{Ab2}
        \label{fig:Ab2}
    \end{subfigure}
    \begin{subfigure}[b]{0.24\textwidth}
        \includegraphics[width=\textwidth]{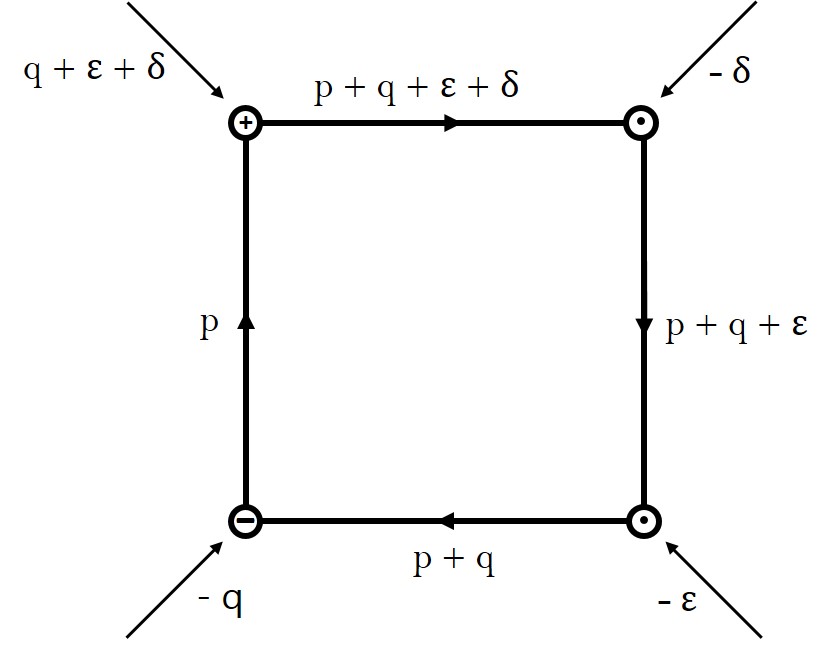}
        \caption{Ac1}
        \label{fig:Ac1}
    \end{subfigure}
    \begin{subfigure}[b]{0.24\textwidth}
        \includegraphics[width=\textwidth]{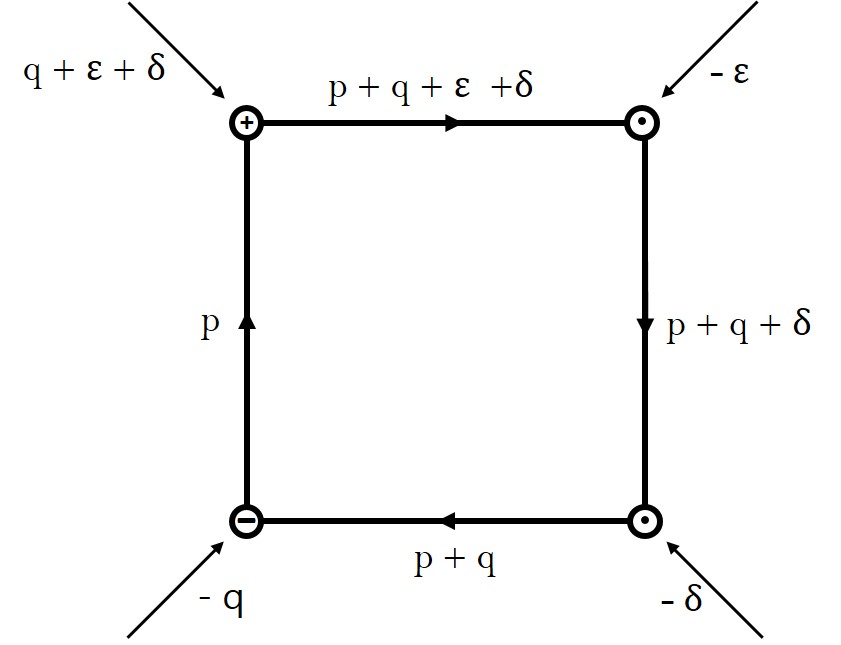}
        \caption{Ac2}
        \label{fig:Ac2}
    \end{subfigure}
    \caption{The box diagrams that contribute to $\langle J_1^{+}(q + \epsilon + \delta) J_1^{-}(-q) J_{0}(-\epsilon) J_{0}(-\delta)\rangle$. All vertices and propagators stand for their exact (all orders in $\lambda$) versions. The vertices are labelled thus -- $\oplus $ for $J_1 ^+ $, $\ominus $ for $J_1 ^- $, and $\odot $ for $J_{0} $. The `A' in the name stands for the fact that these are box diagrams. The second letter denotes the position of the vector vertices (a/b/c for non-adjacent/adjacent horizontal/adjacent vertical ) and changing the number at the end switches the two scalar vertices ($\epsilon \rightarrow \delta $). }\label{fig:box}
\end{figure}

\begin{figure}
    \centering
    \begin{subfigure}[b]{0.24\textwidth}
        \includegraphics[width=\textwidth]{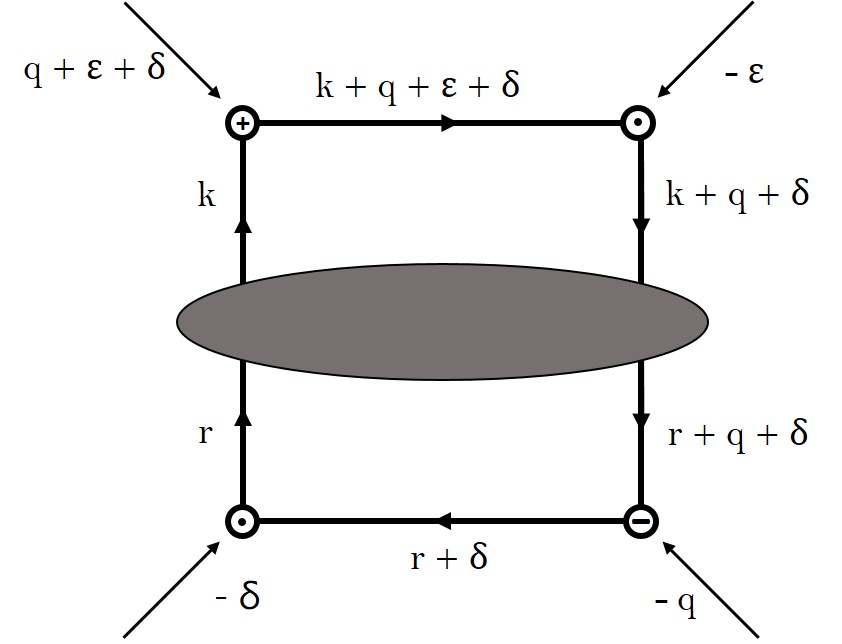}
        \caption{Ba1}
        \label{fig:Ba1}
    \end{subfigure}
    \begin{subfigure}[b]{0.24\textwidth}
        \includegraphics[width=\textwidth]{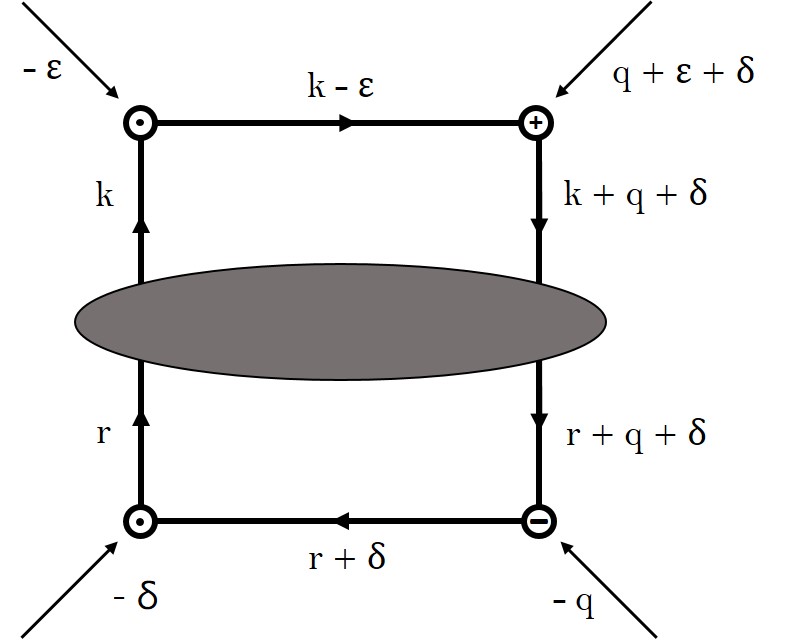}
        \caption{Ba2}
        \label{fig:Ba2}
    \end{subfigure}
    \begin{subfigure}[b]{0.24\textwidth}
        \includegraphics[width=\textwidth]{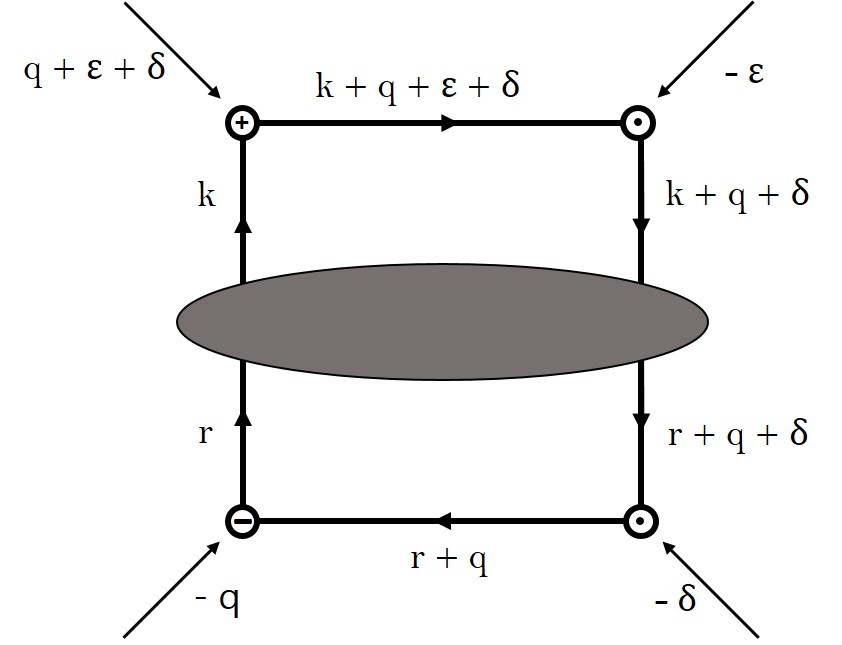}
        \caption{Ba3}
        \label{fig:Ba3}
    \end{subfigure}
    \begin{subfigure}[b]{0.24\textwidth}
        \includegraphics[width=\textwidth]{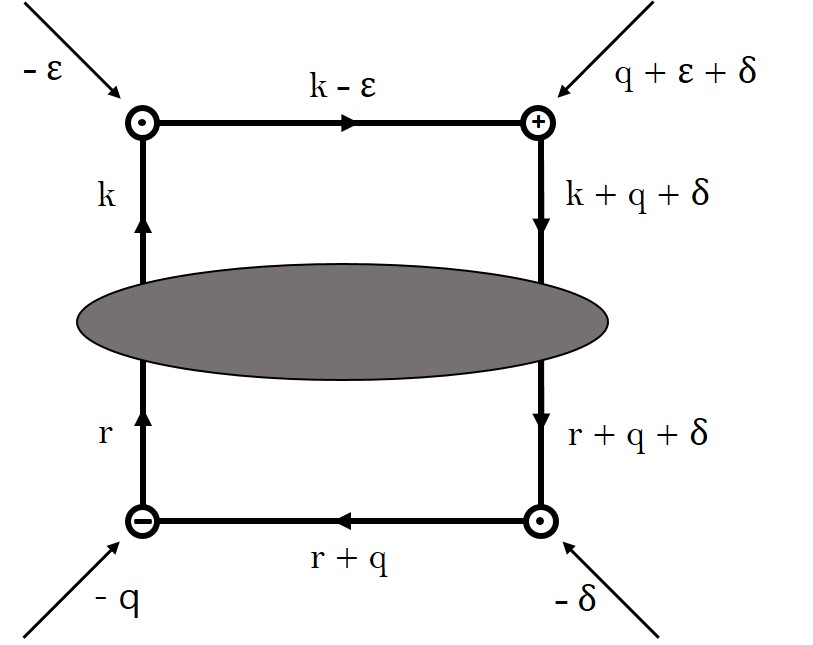}
        \caption{Ba4}
        \label{fig:Ba4}
    \end{subfigure}
    \begin{subfigure}[b]{0.24\textwidth}
        \includegraphics[width=\textwidth]{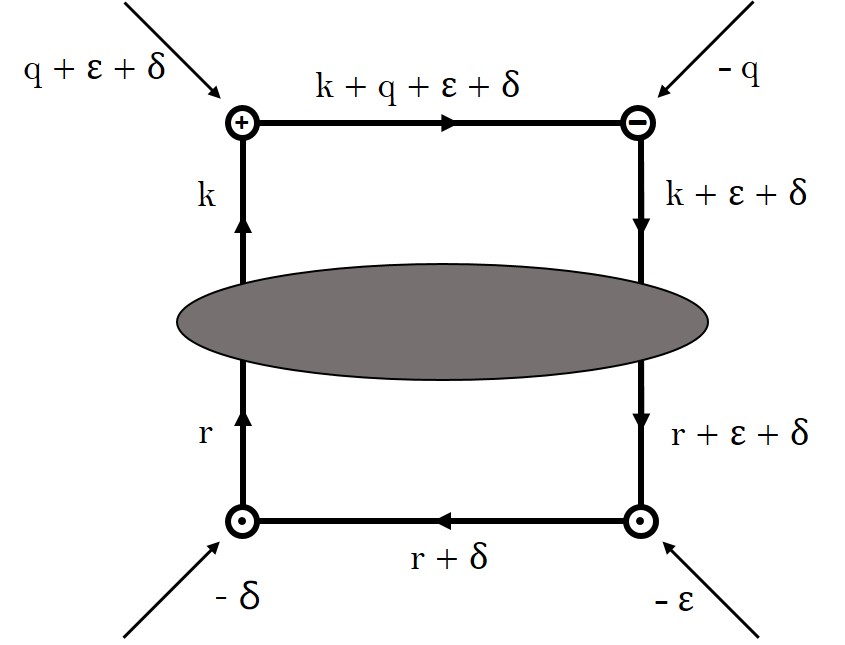}
        \caption{Bb1}
        \label{fig:Bb1}
    \end{subfigure}
    \begin{subfigure}[b]{0.24\textwidth}
        \includegraphics[width=\textwidth]{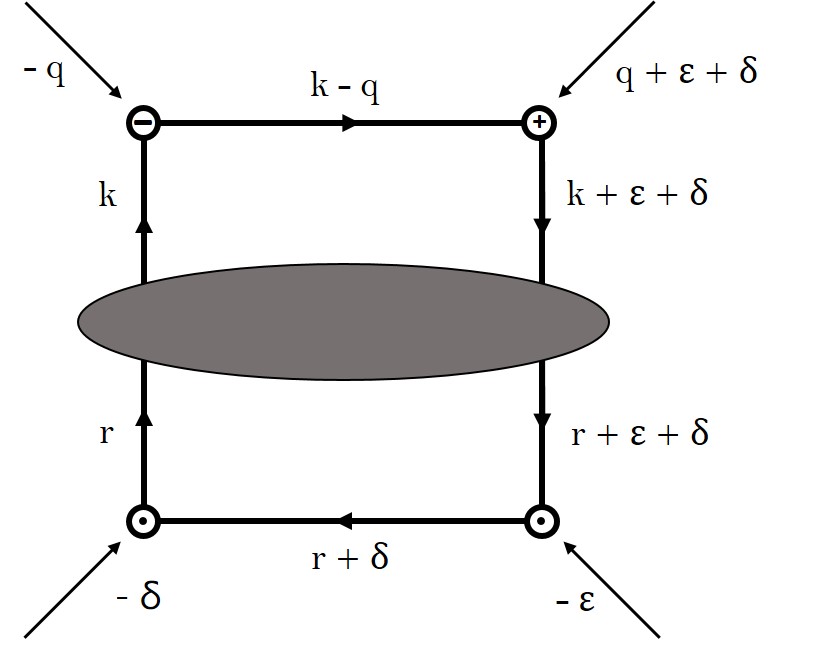}
        \caption{Bb2}
        \label{fig:Bb2}
    \end{subfigure}
    \begin{subfigure}[b]{0.24\textwidth}
        \includegraphics[width=\textwidth]{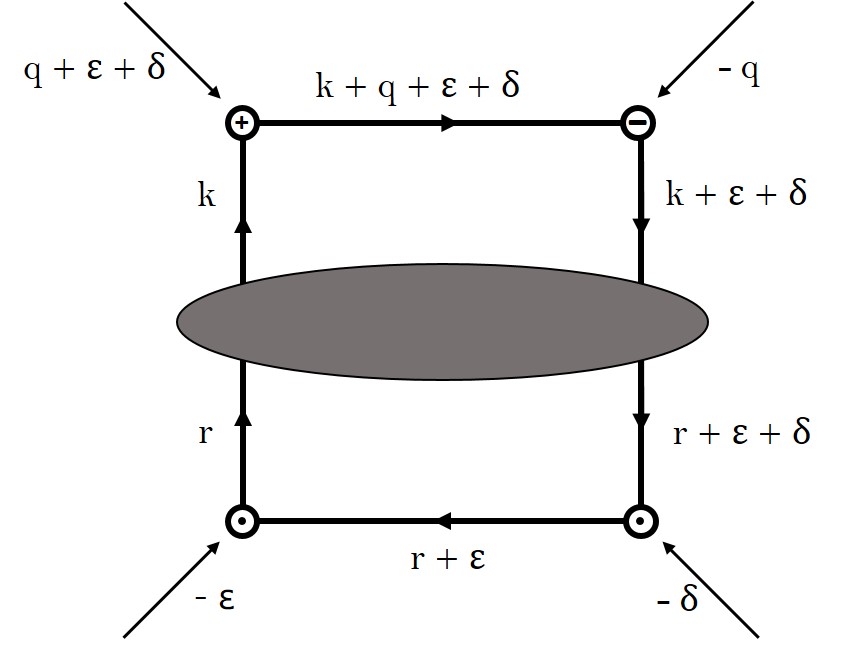}
        \caption{Bb3}
        \label{fig:Bb3}
    \end{subfigure}
    \begin{subfigure}[b]{0.24\textwidth}
        \includegraphics[width=\textwidth]{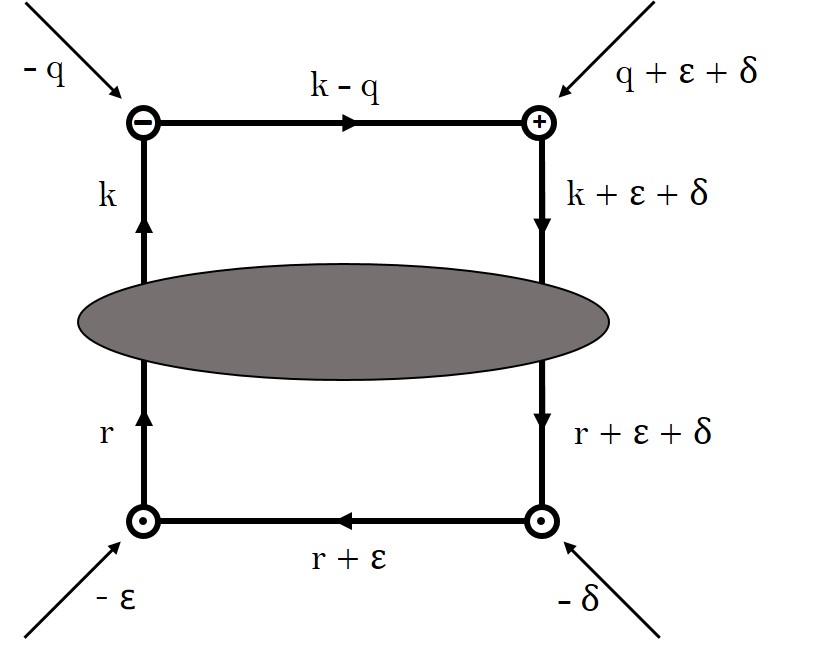}
        \caption{Bb4}
        \label{fig:Bb4}
    \end{subfigure}
    \begin{subfigure}[b]{0.24\textwidth}
        \includegraphics[width=\textwidth]{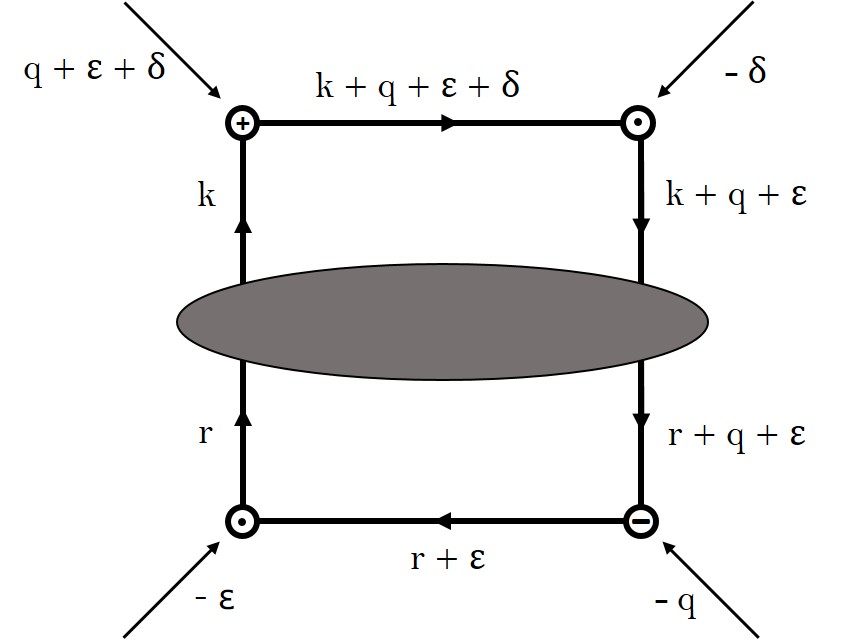}
        \caption{Bc1}
        \label{fig:Bc1}
    \end{subfigure}
    \begin{subfigure}[b]{0.24\textwidth}
        \includegraphics[width=\textwidth]{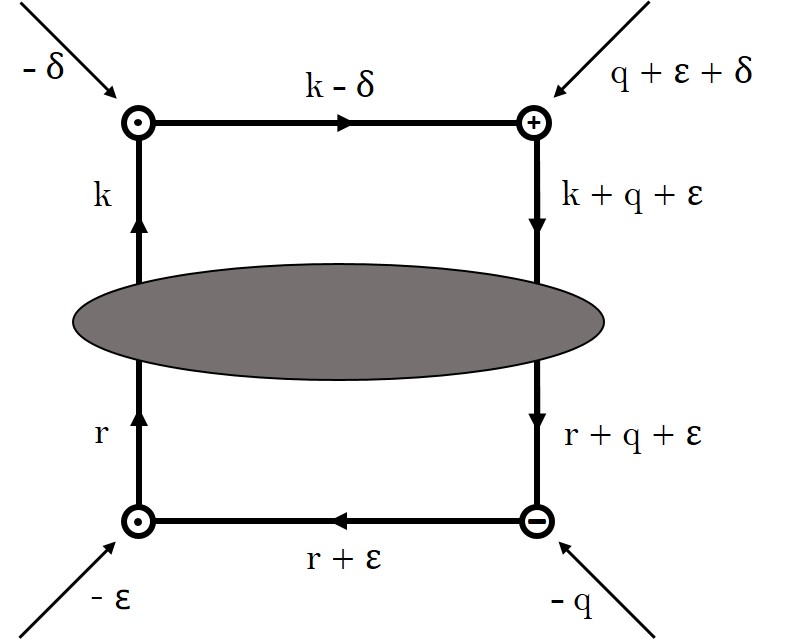}
        \caption{Bc2}
        \label{fig:Bc2}
    \end{subfigure}
    \begin{subfigure}[b]{0.24\textwidth}
        \includegraphics[width=\textwidth]{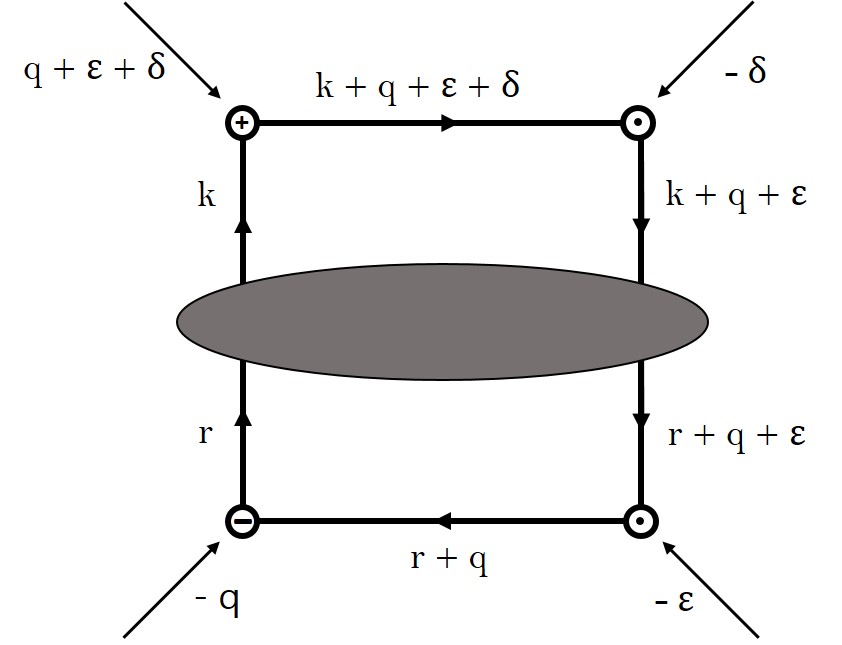}
        \caption{Bc3}
        \label{fig:Bc3}
    \end{subfigure}
    \begin{subfigure}[b]{0.24\textwidth}
        \includegraphics[width=\textwidth]{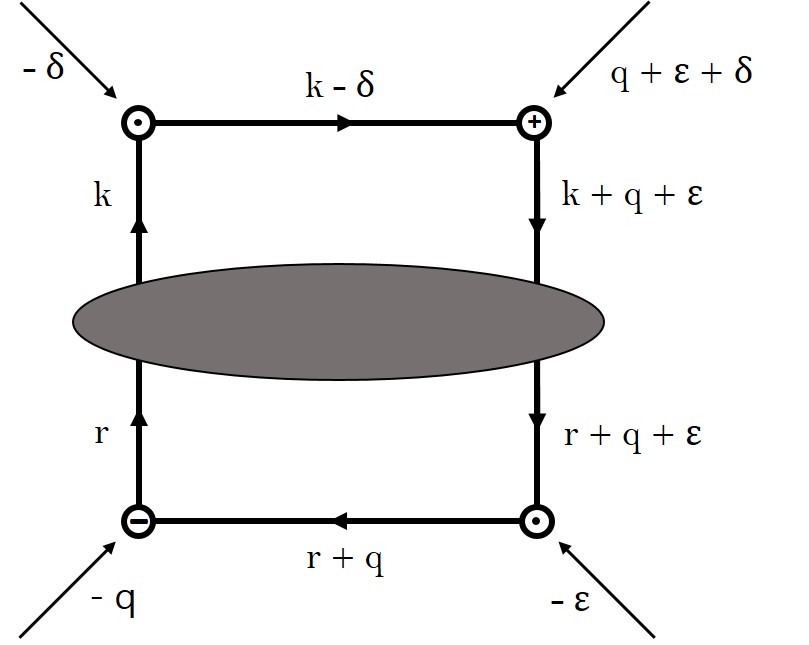}
        \caption{Bc4}
        \label{fig:Bc4}
    \end{subfigure}
    \caption{The ladder diagrams that contribute to $\langle J_1^{+}(q + \epsilon + \delta) J_1^{-}(-q) J_{0}(-\epsilon) J_{0}(-\delta)\rangle$. The `B' in the name stands for the fact that these are ladder diagrams. The second letter stands for the exchange momentum, and changing the number at the end permutes the vertices on each side of the four-point vertex.}\label{fig:lad} 
\end{figure}

The Feynman diagrams that contribute to the correlator may be divided into two main classes -- the box-type (A\#\#) and the ladder-type diagrams (B\#\#)\footnote{The different diagrams in each class are connected by permutations of the insertions.}. Box diagrams are those with no propagators connecting non-adjacent fermion lines, and ladder diagrams those with the effective four-fermion vertex linking non-adjacent fermion lines (see Fig. \ref{fig:box} and \ref{fig:lad}).
We may then use the ingredients discussed earlier to compute these. For example, the generic box diagram is given by: 
\begin{center}
    \centering
    \includegraphics[width=0.24\textwidth]{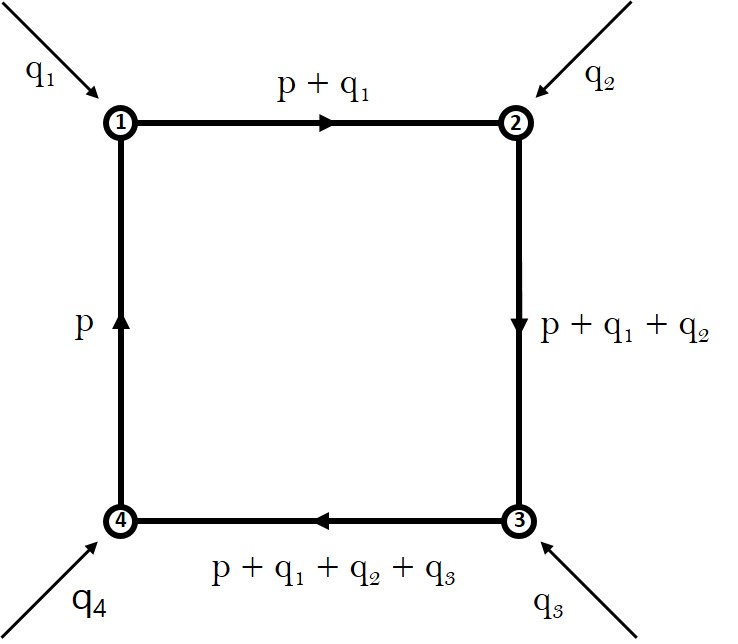},
\end{center}
and corresponds to the expression (we ignore the factors of N for now, and restore them at the end):
\begin{equation}
 - \int \frac{\text{d}^3 p}{(2\pi )^3}  \text{Tr} \left[ \mathcal{O}_1 (q_1) \text{S}(p) \mathcal{O}_4 (q_4) \text{S}(p + q_1 + q_2 + q_3) \mathcal{O}_3 (q_3) \text{S}(p + q_1 + q_2) \mathcal{O}_2 (q_2) \text{S}(p + q_1) \right],
\end{equation}
where the $\mathcal{O}$'s and the S's correspond to the exact vertices and propagators (as defined earlier). Similarly, the ladder diagram:
\begin{center}
    \centering
    \includegraphics[width=0.24\textwidth]{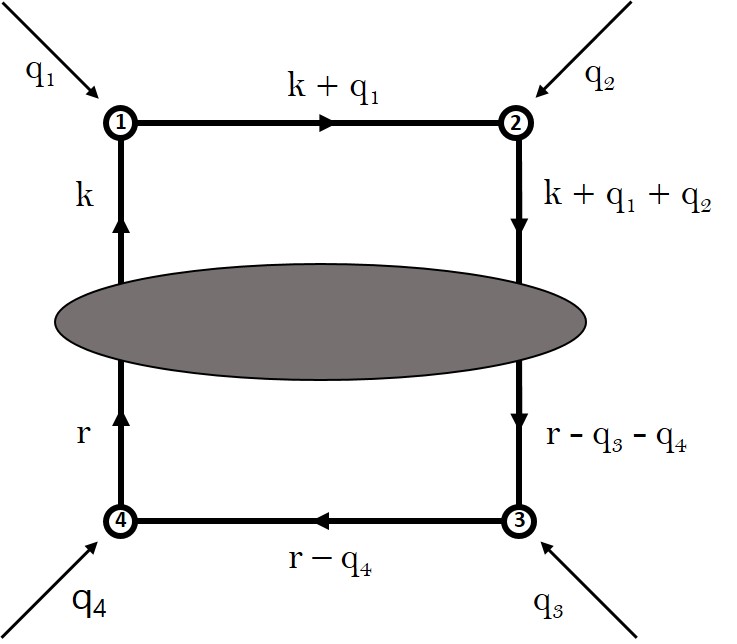},
\end{center}
gives the expression (see Appendix \ref{apd:four_vertex}):
\begin{equation}
\begin{split}
 - \int \frac{\text{d}^3 r}{(2\pi )^3} \frac{\text{d}^3 k}{(2\pi )^3}     \text{Tr} \left[ \text{S}(k + q_1 + q_2) \mathcal{O}_2 \text{S}(k + q_1) \mathcal{O}_1 \text{S}(k)\gamma ^A \text{S}(r) \mathcal{O}_4 \text{S}(r - q_4) \mathcal{O}_3 \text{S}(r - q_3 - q_4) \gamma ^B \right] \\
\Gamma _{AB} (k,q_1 + q_2,r)  .
\end{split}
\end{equation}

\subsection{$p_3$-type integrals}

In each diagram, there are two types of integrals -- the integrals over the third (z) component of the loop momenta or the ``$p_3$-type" integrals, and the integrals over the planar (1 and 2) components or ``$p_s$-type" integrals. We shall discuss these separately.

The box diagrams contain only one $p_3$ integral, while the ladder diagrams contain two. However, these are tackled the same way. We use the fact that the integrands always depend on $p_3$ (or in case of the ladder diagrams, on $r_3$ and $k_3$) in the same way\footnote{The equations that follow are schematic; in particular, $s_i$ are not necessarily the external momenta, and may even be 0.}:

\begin{align}
\begin{split}
\text{Box} \sim {} & \int \text{d}p_3 \frac{1}{((p_3 + s_1)^2 + p_s ^2)((p_3 + s_2)^2 + p_s ^2)((p_3 + s_3)^2 + p_s ^2)((p_3 + s_4)^2 + p_s ^2)}  \\
\text{Ladder} \sim {} & \int \text{d}r_3 \frac{1}{((r_3 + s_1)^2 + r_s ^2)((r_3 + s_2)^2 + r_s ^2)((r_3 + s_3)^2 + r_s ^2)}, \nonumber \\
                   & \ \ \ \ \  \int \text{d}k_3 \frac{1}{((k_3 + \tilde{s}_1)^2 + k_s ^2)((k_3 + \tilde{s}_2)^2 + k_s ^2)((k_3 + \tilde{s}_3)^2 + k_s ^2)}.  \\ 
\end{split} 
\end{align}  
This may be easily seen by noting that the exact vertices under consideration ($J_0$ and $J_1$) and also the ladder, don't depend on the third component of the loop momenta; the only $p_3$ dependence comes from propagator corrections, which take the form $\frac{1}{((p_3 + s)^2 + p_s ^2)}$. These integrals are easy to do.

\subsection{$p_s$-type integrals}

\subsubsection*{Box diagrams}

We shall do the $p_s$ integrals in two steps-- the angular part and the radial part (i. e. , $ \text{d}p_1 \text{d}p_2 = \text{d}\theta _p \ p_s \text{d}p_s$). For the box diagrams, the integrand after $p_3$ integration is independent of $\theta_p$\footnote{This statement applies only to correlators such as $\langle J_1^{+} J_1^{-} J_{0} J_{0}\rangle$, which are invariant to rotations in the 1-2 plane. For other correlators like $\langle J_1^{+} J_{0} J_{0} J_{0}\rangle$, when necessary, we must follow a method similar to the $\text{d} \theta _r $ integration described later on.} and $\int \text{d}\theta _p = 2 \pi$.
The $p_s$ integral, on the other hand, puts up more of a fight. This integral is, in general, intractable \footnote{This is no longer the case in special kinematic regimes such as the double-soft limit, which is what led \cite{Turiaci:2018dht} to consider it.} and we shall set it aside for now...

\subsubsection*{Ladder diagrams}

\paragraph{The angular integrals:}
The ladder diagrams have two loop momenta and hence comprise four integrals -- \newline $\int \text{d}\theta _r \ r_s\text{d}r_s \ \text{d}\theta _k \ k_s\text{d}k_s $. This time, we have a non-trivial angular integral -- $\int \text{d}\theta _r $. To tackle this, we convert the angular integral into a contour integral via the following change of variables:
\begin{equation}
\int _0 ^{2\pi} \text{d}\theta_r = \oint _C \frac{\text{d} r^+ }{i\ r^+} ,
\end{equation}
where `C' stands for the contour $\left| r^+ \right| = \frac{r_s}{\sqrt{2}}$. Hence, doing this integral merely involves computing residues. Examining the ladder and (vector-) vertex dependences on $r^+$ shows that the full diagram generally has the structure $\frac{\left( \text{polynomial in } r^+ \right)}{(r^+ - k^+)r^+ }$. This means that residues can come from two points -- $r^+ = 0$, and also $r^+ = k^+$ depending on if $ \left| r^+ \right| \lessgtr \left| k^+ \right| $ (i.e., $ \left| r_s \right| \lessgtr \left| k_s \right| $). In summary,

\begin{equation}
\int _0 ^{2\pi} \text{d}\theta_r \text{(Ladder)} = \oint _C \frac{\text{d} r^+ }{i\ r^+} \frac{\left( \text{polynomial in } r^+ \right)}{(r^+ - k^+)r^+ } 
	= \left\{ \begin{array}{lcr}
			2 \pi \ \text{Res}_{r^+ = 0}			& , & r_s < k_s \\
				2 \pi \left( \text{Res}_{r^+ = 0} + \text{Res}_{r^+ = k^+} \right) & , & r_s > k_s \\
				\end{array} \right. .
\end{equation}
Once this is done, the other angular integral ($\int \text{d} \theta_k $) becomes trivial\footnote{Again, this statement is true for $\langle J_1^{+} J_1^{-} J_{0} J_{0}\rangle$, but not necessarily for other correlators. In such cases, the method described above is expected to prove useful.} and yields a factor of $2 \pi $.

\paragraph{The planar integrals:}
The first planar integral to be tackled is $\int \text{d}r_s$. Going back to the residue computation, it may be seen that this integral falls into two pieces:

\begin{align}
\int_0 ^{\infty} \text{d}r_s \oint _C \frac{\text{d} r^+ }{i\ r^+} \frac{\left( \text{polynomial in } r^+ \right)}{(r^+ - k^+)r^+ } 
	& = \int_0 ^{\infty} \text{d}r_s \left\{ \begin{array}{lcr}
			2 \pi \ \text{Res}_{r^+ = 0}			& , & r_s < k_s \\
				2 \pi \left( \text{Res}_{r^+ = 0} + \text{Res}_{r^+ = k^+} \right) & , & r_s > k_s \\
				\end{array} \right. \nonumber  \\
				& = \int_0 ^{k_s} 2 \pi \ \text{Res}_{r^+ = 0} + \int_{k_s} ^{\infty} 2 \pi \left( \text{Res}_{r^+ = 0}  + \text{Res}_{r^+ = k^+} \right)  .
\end{align}
Miraculously, these integrals turn out to be soluble\footnote{At least in the case of $\langle J_{0} J_{0} J_{0} J_{0}\rangle$, $\langle J_1^{+} J_1^{-} J_{0} J_{0}\rangle$, and $\langle J_1^{+} J_1^{-} J_1^{+} J_1^{-}\rangle$.}.

However, the residues we get using the method described above are huge, unwieldy expressions. To be able to do anything with these, we have to trim them down to a reasonable form. We choose to do this with the assistance of the various exponentials and their products that pervade our expressions. These come entirely from the exact vertices and the ladder, and are of a certain form -- for example, a diagram from the class \textbf{Bb} contains $e^{ \pm 2i \lambda\ \text{tan}^{-1}\left( \frac{2 r_s}{\delta} \right) } $, $e^{ \pm 2i \lambda\ \text{tan}^{-1}\left( \frac{2 r_s}{\epsilon} \right) } $ (from the vertices in the upper half -- hence the loop momenta and external momenta are from the upper half), $e^{ \pm 2i \lambda\ \text{tan}^{-1}\left( \frac{2 k_s}{q + \delta + \epsilon} \right) } $, $e^{ \pm 2i \lambda\ \text{tan}^{-1}\left( \frac{2 k_s}{q} \right) } $ (from the vertices in the lower half), $e^{ \pm 2i \lambda\ \text{tan}^{-1}\left( \frac{2 r_s}{\delta + \epsilon} \right) } $, $e^{ \pm 2i \lambda\ \text{tan}^{-1}\left( \frac{2 k_s}{\delta + \epsilon} \right) } $ (from the ladder -- the momentum that appears is the momentum exchanged between the upper and lower halves in this subclass) and their products. Rather than simplifying the whole expression at once, we treat the full expression as a polynomial in these exponentials, extract their coefficients (taken individually, these contain fewer terms), and simplify then separately. Once the dust has cleared, we find that the $r_s$ integrals may be done as discussed earlier.
 
But the $k_s$ integrals, like the $p_s$ integrals, are not soluble via Mathematica. So we once again, store these expressions and set them aside...

\subsection{Arriving at the result}

After the procedure described above is complete, we are left with 18 integrals that cannot be done analytically, at least via Mathematica. However, a numerical evaluation will quickly reveal that all of these integrals are (at least in general) convergent. This lets us employ the following trick:
%(in our notebooks, we sum diagrams in certain subclasses, so the number of integrals comes down to 9, but the problem remains)

\begin{equation}
\int_0 ^{\Lambda} \text{d}p_s\ \text{Box}(p_s) + \int_0 ^{\Lambda} \text{d}k_s\ \text{Ladder}(k_s) = \int_0 ^{\Lambda} \text{d}p_s\ \left( \text{Box}(p_s) + \text{Ladder}(p_s) \right)
\end{equation}
(of course, the equation above is schematic; we have more than one box and ladder integral). The integral created thus turns out to be soluble and gives us a cut-off dependent expression for the correlator. The final results (listed in Section \ref{sec:res}) are then obtained by sending the cut-off to infinity.

% Section 4---------------------------------------------------------------------------

\section{Results and discussion}
\label{sec:res}

\subsection{$\lambda $ dependence via bootstrap arguments}

Let's first look at what we can say about the correlators from bootstrap arguments, and the bosonisation duality. The conformal block expansion (in any channel) for $\langle J_0 J_0 J_0 J_0 \rangle$ may be written schematically as:
\begin{IEEEeqnarray}{rCl}
\langle \tilde{J}_0 \tilde{J}_0 \tilde{J}_0 \tilde{J}_0 \rangle & \sim & \sum_{\text{st}}^{} \langle \tilde{J}_0 \tilde{J}_0 \vert \tilde{J}_s \rangle \langle \tilde{J}_s \vert \tilde{J}_0 \tilde{J}_0 \rangle 
+ \sum_{\mathbf{1}, \text{dt}}^{} \langle \tilde{J}_0 \tilde{J}_0 \vert \mathcal{O} \rangle \langle \mathcal{O} \vert \tilde{J}_0 \tilde{J}_0 \rangle + \mathcal{O} \left( \frac{1}{N^2} \right)
\end{IEEEeqnarray}
where the $\mathbf{1}$, `st' and `dt' stand for the identity operator, single-trace operators, and double-trace operators respectively. Note that we've used the appropriately normalised version of the single-trace operators:
\begin{subequations}
\begin{align}
\tilde{J}_0 & = \frac{1}{\sqrt{\tilde{N}}\left( 1 + \tilde{\lambda} ^2 \right)  } J_0, \\
\tilde{J}_s & = \frac{1}{\sqrt{\tilde{N}} } J_s.
\end{align}
\end{subequations}
where
\begin{equation}
\tilde{N} = \frac{2\ N\ \text{sin}\left( \pi \lambda \right)}{\pi \lambda },
\end{equation}
\begin{equation}
\tilde{\lambda} = \text{tan}\left( \frac{\pi \lambda }{2} \right).
\end{equation}

As shown in \cite{Heemskerk:2009pn}, the three-point function $\langle J_0 J_0 \mathcal{O}_{\text{dt}} \rangle $ at $\mathcal{O}( 1 / \tilde{N} )$ (it starts at $ \mathcal{O} ( \tilde{N}^0 ) $ in our normalisation) receives corrections from two sources -- the OPE coefficient $c_{0 0 \mathcal{O}}^{(1)}$, and the anomalous dimension $\gamma^{(1)}_{\mathcal{O}}$; hence, the crossing equation may be written as an inhomogeneous, linear equation (actually, an infinite number of linear equations, since both sides are functions) for these quantities, sourced by the single-trace contributions:
\begin{IEEEeqnarray}{rCl}
\sum_{\text{dt}}^{} \left( \frac{ 1 }{ \tilde{N} }\gamma^{(1)}_{\mathcal{O}}  \ \#  + \frac{1}{\tilde{N}} c_{0 0 \mathcal{O}}^{(1)} \ \# \right) & \sim & \sum_{\text{st}}^{} \frac{1}{ \tilde{N}} \frac{ 1 }{ ( 1 + \tilde{ \lambda }^2 )^2 } \#
\label{eqn:cross}
\end{IEEEeqnarray}
Here, we use the known form of the single-trace three point functions at $\mathcal{O} \left( 1 / \sqrt{\tilde{N}} \right) $ \cite{Maldacena:2012sf}. What we'll be interested in are their $\tilde{ \lambda }$-dependences:
\begin{IEEEeqnarray}{rCl}
\langle \tilde{J}_0 \tilde{J}_0 \tilde{J}_s \rangle & \sim & \frac{1}{ \sqrt{ \tilde{N}} } \frac{ 1 }{  1 + \tilde{ \lambda }^2  } \# \\
\langle \tilde{J}_0 \tilde{J}_1 \tilde{J}_s \rangle & \sim & \frac{1}{ \sqrt{ \tilde{N}} } \frac{ 1 }{  1 + \tilde{ \lambda }^2  } \# + 
\frac{1}{ \sqrt{ \tilde{N}} } \frac{ \tilde{ \lambda } }{  1 + \tilde{ \lambda }^2  } \# \\
\langle \tilde{J}_1 \tilde{J}_1 \tilde{J}_s \rangle & \sim & \frac{1}{ \sqrt{ \tilde{N}} } \frac{ 1 }{  1 + \tilde{ \lambda }^2  } \# + 
\frac{1}{ \sqrt{ \tilde{N}} } \frac{ \tilde{ \lambda } }{  1 + \tilde{ \lambda }^2  } \#  + \frac{1}{ \sqrt{ \tilde{N}} } \frac{ \tilde{ \lambda }^2 }{  1 + \tilde{ \lambda }^2  } \#
\end{IEEEeqnarray}
``Solving'' for the four-point function at $\mathcal{O} ( 1 / \tilde{N} ) $ means finding the $ \gamma^{(1)}_{\mathcal{O}} $'s and the $ c_{0 0 \mathcal{O}}^{(1)} $'s (not that we'll be doing either of those). 

Now we observe that sending $\tilde{ \lambda } $ to zero means sending the right-hand side of (\ref{eqn:cross}) to that of the free (fermionic) theory, which is solved by the corresponding parameters of the free theory. This means that a possible solution to (\ref{eqn:cross}) is $ \left( \frac{ \gamma^{(1)}_{\mathcal{O}; \text{fer}} }{ ( 1 + \tilde{\lambda } ^2 )^2 } , \frac{ c_{0 0 \mathcal{O}; \text{fer}}^{(1)} }{{ ( 1 + \tilde{\lambda } ^2 )^2 } } \right) $; this produces a four-point correlator proportional to that of the free theory\footnote{In the free theory, all the anomalous dimensions $ \gamma^{(1)}_{\mathcal{O}; \text{fer}} $ vanish, but this doesn't affect the argument.}.  However, we may add to this any solution of the corresponding homogeneous equation (i.e., with 0 on the right-hand side of (\ref{eqn:cross})). In fact, these solutions were analysed by \cite{Heemskerk:2009pn} and correspond to contact Witten diagrams in $AdS$.

It was then discovered by \cite{Turiaci:2018dht} that only three out of the infinitely many of these ``truncated" solutions may be added to the inhomogeneous solution without spoiling the Regge behaviour of the correlator. The most general solution to crossing which is well-behaved in the Regge limit is then:
\begin{IEEEeqnarray}{rCl}
\langle \tilde{J}_0 \tilde{J}_0 \tilde{J}_0 \tilde{J}_0 \rangle ^{\text{conn.}} & = & \frac{1}{ \tilde{N} } \left( \frac{1}{ ( 1 + \tilde{ \lambda }^2 )^2 } \langle J_0 J_0 J_0 J_0 \rangle _{ \text{free}} + b_1 G^{AdS}_{\phi^4} +  b_2 G^{AdS}_{ \left( \partial\phi \right)^4} + b_3 G^{AdS}_{\phi^2 \left( \partial ^3 \phi \right)^2} \right) + \mathcal{O} \left( \frac{1}{ \tilde{N}^2 } \right) \nonumber \\
\end{IEEEeqnarray}
where $G^{AdS}_{\mathcal{V}} $ is the contact diagram with vertex $\mathcal{V}$. Furthermore, an explicit numerical evaluation of the correlator by summing the planar diagrams \cite{Turiaci:2018dht} (alternatively, the explicit result that we obtained above) showed that $b_1 = b_2 = b_3 = 0 $ and that the full correlator at $\mathcal{O} ( 1 / \tilde{N} )$ was indeed proportional to the free theory answer. We note that the same result was obtained recently by \cite{Li:2019twz} via an explicit solution of the partial higher-spin Ward identities, which means that modifications by truncated solutions are simply incompatible with higher-spin symmetry. The resultant object also satisfies the criterion of analyticity in spin. It seems that in the case of $\langle J_0 J_0 J_0 J_0 \rangle$ (and also $  \langle J_2 J_0 J_0 J_0 \rangle $ as shown by \cite{Li:2019twz}), weakly broken higher-spin symmetry culls any contact terms\footnote{We use the `contact terms' to mean terms that arise from contact diagrams in $AdS_4$. They are not to be confused with delta functions in position space.} that spoil analyticity in spin. 

Similar arguments go thorough for the other correlators under our consideration -- $\langle J_1 J_0 J_1 J_0 \rangle $ and $ \langle J_1 J_1 J_1 J_1 \rangle $. This time though, the three-point functions have more $\tilde{ \lambda } $-structures, leading to: 
\begin{IEEEeqnarray}{rCl}
\sum_{\text{st}} \langle \tilde{J}_1 \tilde{J}_0 \tilde{J}_s \rangle \langle \tilde{J}_s \tilde{J}_1 \tilde{J}_0 \rangle & \sim & \frac{1}{ \tilde{N} } \frac{ 1 }{ ( 1 + \tilde{ \lambda }^2 )^2 } \# + 
\frac{1}{  \tilde{N} } \frac{ \tilde{ \lambda } }{ ( 1 + \tilde{ \lambda }^2 )^2 }  \# +
\frac{1}{ \tilde{N} } \frac{ \tilde{ \lambda }^2 }{ ( 1 + \tilde{ \lambda }^2 )^2 }  \# \\
\sum_{\text{st}} \langle \tilde{J}_1 \tilde{J}_1 \tilde{J}_s \rangle \langle \tilde{J}_s \tilde{J}_1 \tilde{J}_1 \rangle & \sim & \frac{1}{ \tilde{N} } \frac{ 1 }{ ( 1 + \tilde{ \lambda }^2 )^2 } \# + 
\frac{1}{  \tilde{N} } \frac{ \tilde{ \lambda } }{ ( 1 + \tilde{ \lambda }^2 )^2 }  \# + 
\frac{1}{  \tilde{N} } \frac{ \tilde{ \lambda }^2 }{ ( 1 + \tilde{ \lambda }^2 )^2 }  \#
+
\frac{1}{ \tilde{N} } \frac{ \tilde{ \lambda }^3 }{ ( 1 + \tilde{ \lambda }^2 )^2 }  \# \nonumber \\
& & \ \ \ \ \ \ \ \ \ \ \ \ \ \ \ \ \ \ \ \ \ \ \ \ \ \ \ \ \ \ \ \ \ \ \ \ \ \ \ \ \ \ \ \ \ \ \ \ \ \ \ \ \ \ \ +
\frac{1}{ \tilde{N} } \frac{ \tilde{ \lambda }^4 }{ ( 1 + \tilde{ \lambda }^2 )^2 }  \# 
\end{IEEEeqnarray}
Of course, we are playing with mixed, spinning correlators and hence have to deal with the complications that come with this (see \cite{Kos:2014bka} for details), but all that we need to know here is that in the end, the right-hand sides of the crossing equations (in case of $\langle J_1 J_0 J_1 J_0 \rangle$, the one with the correct channels) have 3 and 5 $ \tilde{ \lambda } $-structures respectively, and the analogue of (\ref{eqn:cross}) can be solved separately for each one. If this could be done, the resultant solutions would look like:
\begin{align}
\langle \tilde{J}_1 \tilde{J}_0 \tilde{J}_1 \tilde{J}_0 \rangle ^{\text{conn.}} & \sim  \frac{1}{\tilde{N}} \left(  \frac{1}{(1 + \tilde{\lambda}^{2} )^2 }  \langle \tilde{J}_1 \tilde{J}_0 \tilde{J}_1 \tilde{J}_0 \rangle ^{\text{conn.}}_{\text{fer}} +  \frac{\tilde{\lambda}}{( 1 + \tilde{\lambda}^{2} )^2 }  (\#) +  \frac{\tilde{\lambda}^2 }{( 1 + \tilde{\lambda}^{2} )^2 }  \langle \tilde{J}_1 \tilde{J}_0 \tilde{J}_1 \tilde{J}_0 \rangle ^{\text{conn.}}_{\text{bos}} \right) \label{eqn:corr1}
\\
\langle \tilde{J}_1 \tilde{J}_1 \tilde{J}_1 \tilde{J}_1 \rangle ^{\text{conn.}} & \sim  \frac{1}{\tilde{N}} \left(  
\frac{1}{(1 + \tilde{\lambda}^{2} )^2 } \langle \tilde{J}_1 \tilde{J}_1 \tilde{J}_1 \tilde{J}_1 \rangle ^{\text{conn.}}_{\text{fer}} 
+  \frac{\tilde{\lambda}}{( 1 + \tilde{\lambda}^{2} )^2 }  (\#) 
+  \frac{\tilde{\lambda}^2 }{( 1 + \tilde{\lambda}^{2} )^2 }  (\#) 
+ \frac{\tilde{\lambda}^3 }{( 1 + \tilde{\lambda}^{2} )^2 } (\#) \right.
\nonumber \\
& \ \ \ \ \ \ \ \ \ \ \ \ \ \ \ \ \ \ \ \ \ \ \ \ \ \ \ \ \ \ \ \ \ \ \ \ \ \ \ \ \ \ \ \ \ \ \ \ \ \left. + \frac{\tilde{\lambda}^4 }{( 1 + \tilde{\lambda}^{2} )^2 } \langle \tilde{J}_1 \tilde{J}_1 \tilde{J}_1 \tilde{J}_1 \rangle ^{\text{conn.}}_{\text{bos}}
\right)
\label{eqn:corr2}
\end{align}
This will be the form that we expect our solutions to take. Note that $ \tilde{ \lambda } \rightarrow 0, \infty $ turns the crossing equations into those of the free fermionic and critical scalar respectively, and hence in these limits, we expect the correlator to go over to those theories\footnote{Recall that the scalar operator in the critical theory is given by $J_0 ^{\text{cb}} := \tilde{\lambda} J_0 $ and hence $ \langle J_1 ^+ J_1 ^- J_0 ^{\text{cb}} J_0 ^{\text{cb}} \rangle = \lim _{\tilde{\lambda} \rightarrow \infty} \langle J_1 ^+ J_1 ^- \left( \tilde{\lambda} J_0 \right) \left( \tilde{\lambda} J_0 \right) \rangle $ is finite.}.

Once again, any homogeneous solution may be added to these solutions with arbitrary $ \tilde{ \lambda } $ dependence as long as they don't contribute in the $ \tilde{ \lambda } \rightarrow 0, \infty $ limit. However, we have verified that possible homogeneous solutions give non-vanishing answers in the collinear limit, and hence any new $ \tilde{ \lambda } $ dependences would be represented in the collinear limit. In the following, we will find that our results do not exhibit such dependences and are in compliance with the conjecture that $ \langle J_0 J_0 J_1 J_1 \rangle $ and $ \langle J_1 J_1 J_1 J_1 \rangle $ depend on $ \tilde{ \lambda } $ exactly as in Eqn. (\ref{eqn:corr1}) and (\ref{eqn:corr2}).

\subsection{Results}

We list our main results from the computations of the previous section, in the parameterisation of Maldacena and Zhiboedov\cite{Maldacena:2012sf}. We also recall that all external momenta are along the z-axis, and that
\begin{equation}
\langle ... \rangle = \langle \langle ... \rangle \rangle \ ( 2 \pi )^3 \delta^3 \left( \sum_i q_i \right).
\end{equation} 

\subsubsection{$\left< J_0 J_0 J_0 J_0 \right> $}

\begin{align}
\langle \langle \tilde{J_{0}}(p) \tilde{J_{0}}(q)\tilde{J_{0}}(\epsilon) \tilde{J_{0}} (\delta) \rangle \rangle & =  \frac{1}{\tilde{N} (1 +  \tilde{\lambda }^2 )^2}  \left[ \frac{-\left( p \left| p \right| + q \left| q \right| + \epsilon \left| \epsilon \right| + \delta \left| \delta \right| \right)}{4\left( p + q \right) \left( p + \epsilon \right) \left( p + \delta \right)} \right] 
\end{align}
This matches the result obtained in \cite{Turiaci:2018dht} up to constant factors. The result is proportional to the free theory as was discussed above.

\subsubsection{$\left< J_1 J_1 J_0 J_0 \right> $}

\begin{gather}
\langle \langle \tilde{J}_1^{+}(p) \tilde{ J }_1^{-}(q)\tilde{J_{0}}(\epsilon) \tilde{J_{0}} (\delta) \rangle \rangle  =  \frac{1}{\tilde{N} (1 +  \tilde{\lambda }^2 )^2} 
 \left[ -\frac{1}{4} \frac{q \left| p \right| + p \left| q \right| + \epsilon  \left| \delta \right| + \delta \left| \epsilon \right| }{\left(\epsilon + \delta \right) \left(q + \epsilon \right) \left(q + \delta \right)}  -
 \frac{1}{4} \frac{- \left| \delta \right| - \left| \epsilon \right| +  \left| q + \epsilon \right| + \left| q + \delta \right| }{q p}  \right]
 \nonumber \\
 +
  \frac{\tilde{\lambda }}{\tilde{N} (1 +  \tilde{\lambda }^2 )^2} 
  \left[
  -\iu \frac{(p - q)}{2 p q} - \frac{\iu}{8} \left( \frac{\left( \left| q \right| q + \left| p \right| p \right) \left( p ^2 - q ^2\right) + 
   \left( \epsilon ^2 + \delta ^2 \right) \left( \left| q \right| q - \left| p \right| p \right) }{p q \left( p + \delta \right) \left( p + \epsilon \right) \left( p + q \right)}  \right) \left( \text{sgn}(\epsilon) + \text{sgn}(\delta) \right)  
\right. \nonumber
 \\
\left.
- \iu \frac{ \left| p + \epsilon \right| - \left| p + \delta \right| }{4 p q } \left( \text{sgn}(\delta) - \text{sgn} (\epsilon) \right)
\right]  
 \nonumber \\
   + \frac{\tilde{\lambda 
}^2}{\tilde{N}(1 +  \tilde{\lambda }^2 )^2} 
\left[
-\frac{ \left( q \left| p \right| + p \left| q \right| \right) \text{sgn}(\epsilon) \text{sgn}(\delta) + \epsilon \left| \delta \right| + \delta \left| \epsilon \right| }{4    \left( p + \delta \right) \left( p + \epsilon \right) \left( p + q \right)}
-
\frac{ \left| \epsilon \right| \left| \delta \right| \left( \left| p + \epsilon \right| + \left| p + \delta \right|  \right) + \left( \left| \epsilon \right| + \left| \delta \right| \right) \epsilon \delta  }{ 4 p q \epsilon \delta }
\right.
 \nonumber \\
\left.
- \frac{ \left( q \left| q \right| + p \left| p \right| \right) \left( 1 + \text{sgn}(\epsilon)\text{sgn}(\delta) \right) }{4p q \left( \epsilon + \delta \right) } \right] 
\end{gather}
We see that this is pretty much what we expected. The first and last structures may be matched with the four-point functions from the relevant limiting theories.

\subsubsection{ $\left< J_1 J_1 J_1 J_1 \right> $}

The full expression  for $\left< J_1 J_1 J_1 J_1 \right>$ is too unwieldy to put here, and is relegated to the appendix. We quote here the result for the special case where $q  >\epsilon > \delta >0 $:
\begin{gather}
\langle \tilde{J}_1^{+}(q + \epsilon + \delta) \tilde{J}_1^{-}(-q) \tilde{J}_1^{+}(-\epsilon) \tilde{J}_1^{-} (-\delta) \rangle = \frac{1}{\tilde{N}(1 +  \tilde{\lambda }^2 )^2} \ \left( -\frac{\delta q \epsilon}{4(\delta + q)(q + \epsilon)(\delta + \epsilon)(q + \delta + \epsilon)} \right)  \nonumber \\
 + \frac{ \tilde{\lambda }}{\tilde{N} (1 +  \tilde{\lambda }^2 )^2} \ \left( -\frac{\iu \delta q \epsilon}{2(\delta + q)(q + \epsilon)(\delta + \epsilon)(q + \delta + \epsilon)}  \right)
 + \frac{ \tilde{\lambda }^3}{\tilde{N} (1 +  \tilde{\lambda }^2 )^2} \ \left( -\frac{\iu \delta q \epsilon}{2(\delta + q)(q + \epsilon)(\delta + \epsilon)(q + \delta + \epsilon)}  \right) \nonumber  \\
 + \frac{\tilde{\lambda }^4}{\tilde{N}(1 +  \tilde{\lambda }^2 )^2} \ \left(  \frac{\delta q \epsilon}{4(\delta + q)(q + \epsilon)(\delta + \epsilon)(q + \delta + \epsilon)}  \right) \nonumber  \\ 
=  \frac{1}{\tilde{N}(1 +  \tilde{\lambda }^2 )} \ \left(  -\frac{\delta q \epsilon}{4(\delta + q)(q + \epsilon)(\delta + \epsilon)(q + \delta + \epsilon)} \right) 
 + \frac{ \tilde{\lambda }}{\tilde{N}(1 +  \tilde{\lambda }^2 )} \ \left(  -\frac{\iu \delta q \epsilon}{2(\delta + q)(q + \epsilon)(\delta + \epsilon)(q + \delta + \epsilon)}  \right) \nonumber  \\
 + \frac{\tilde{\lambda }^2}{\tilde{N}(1 +  \tilde{\lambda }^2 )} \ \left(  \frac{\delta q \epsilon}{4(\delta + q)(q + \epsilon)(\delta + \epsilon)(q + \delta + \epsilon)} \right) \nonumber \\
 \label{eqn:cancel}
 =  \frac{ ( \tilde{\lambda } - \iu )^2}{\tilde{N}(1 +  \tilde{\lambda }^2 )} \ \left(  \frac{\delta q \epsilon}{4(\delta + q)(q + \epsilon)(\delta + \epsilon)(q + \delta + \epsilon)} \right).
 \end{gather}

Looking at the result, we find a curious coincidence: the tensor structures cancel among themselves to leave three pieces, each with a factor of $\frac{ \tilde{\lambda} ^i }{ 1 + \tilde{\lambda}^2 } $, $ i \in \{ 0, ..., 2 \}$. This is the second equality in \eqref{eqn:cancel}, and it also holds for the general expression; the third equality (the relation between the three different structures) doesn't happen in the general case. In any case, this expression (and also the one in the appendix) is in line with our expectations, up to possible terms that vanish in the collinear limit.

As was discussed before, we may add to the expected form (\ref{eqn:corr2}) of the solution, any homogeneous solution to the crossing equation with an arbitrary $\lambda$-dependence; and still satisfy crossing. What we've done here is to verify that $\langle J_1 J_1 J_1 J_1 \rangle$ in the collinear limit has the expected $\lambda $-dependence. This means that any homogeneous solution that does not vanish in the collinear limit cannot come with an arbitrary $\lambda$ dependence. In other words, the full correlator has the expected $\lambda$-dependence up to those homogeneous solutions, which are well-behaved in the Regge limit, and vanish in the collinear limit.

Like in the scalar case, there are a finite number of candidate homogeneous solutions that do not spoil the Regge behaviour of the correlator. In the $ \langle J_1 J_1 J_1 J_1 \rangle $ case,  these are\footnote{We thank S. Minwalla for notifying us of this result of \cite{Minwalla:contact}.} the 4-pt Witten diagrams in $AdS_4$, with four external gauge fields, and with vertices $ (F^2)^2 $, $ F^4 $, and $ F^{\mu \nu } \left( \partial _{\mu}  F_{\rho} ^{\sigma} \right) \left( \partial _{\nu} F_{\sigma} ^{\lambda } \right) F_{\lambda} ^{\rho}$. We have determined that these diagrams give non-zero answers in the collinear limit; this means that these solutions cannot contribute to the general (with arbitrary external momenta) answer with arbitrary $ \tilde{ \lambda } $ dependences, while still being compliant with our results. This indicates that the full correlator has the same $ \tilde{ \lambda } $ dependence as the collinear one.

In the $ \langle J_0 J_0 J_1 J_1 \rangle $ case, we expect that the contributing vertices will have two scalar fields, two field strength tensors, and up to two derivatives ($ \phi^2 F \wedge F $ and  $ \phi^2 \mathcal{D}_{\mu} F_{\rho \sigma} \mathcal{D}^{\mu} F^{\rho \sigma} $ for instance), though this has not been proven as far as we know. We do not know exactly which of these diagrams are relevant, but have verified that any such vertex (up to total derivatives) gives a non-vanishing contribution in the collinear limit, so we can reach the same conclusion about the $\tilde{ \lambda } $-dependence.

\subsubsection{$\left<  J_1 ^3 ...  \right> $}

When $J_1^{3}$ is introduced into the mix (this is the z-component of the spin-1 current), we find that the correlator vanishes (in the collinear limit, that is). For example:
\begin{subequations}
\begin{align}
\langle J_1^{3}(q + \epsilon + \delta) J_1^{3}(-q)J_1^{3}(-\epsilon) J_1^{3} (-\delta) \rangle & =  0  \\
\langle J_1^{3}(q + \epsilon + \delta) J_1^{3}(-q) J_{0}(-\epsilon) J_{0} (-\delta) \rangle & =  0  \\
\langle J_1^{3}(q + \epsilon + \delta) J_{0}(-q) J_{0}(-\epsilon) J_{0} (-\delta) \rangle & =  0  \\
\langle J_1^{3}(q + \epsilon + \delta) J_1^{3}(-q)J_1^{+}(-\epsilon)J_1^{-} (-\delta) \rangle & =  0  \\
\langle J_1^{3}(q + \epsilon + \delta) J_{0}(-q) J_1^{+}(-\epsilon) J_1^{-} (-\delta) \rangle & =  0  .
\end{align}
\end{subequations}
This is not very surprising and is a consequence of conservation of $J_{1}$ and the fact that all the $J$'s are neutral. We emphasize that that this does not mean that the correlators themselves vanish. For example, we have verified by explicit computation that $\langle J_1^{3}(q + \epsilon + \delta) J_{0}(-q)J_{0}(-\epsilon) J_{0} (-\delta) \rangle$ does not vanish for certain non-collinear configurations. More powerful methods are needed to compute these correlators for all $\lambda$.
 
  \subsection{Conclusions and future directions}
 
 We have computed, in the collinear limit, and at leading order in $1/N$, all the four-point functions that involve $J_0$ and $J_1$ in CS fermionic theories\footnote{We haven't looked at $\langle J_1^+ J_1^+ J_0 J_0 \rangle$, etc which are charged under rotations in the x-y plane, and thus vanish when all momenta are in the $z$ direction.}. Specifically, we have analytic expressions for $\langle J_0 J_0 J_0 J_0 \rangle$, which is in agreement with the result of \cite{Turiaci:2018dht}, and for $\langle J_1^+ J_1^- J_0 J_0 \rangle$ and $\langle J_1^+ J_1^- J_1^+ J_1^- \rangle$, that fall in line with expectations from bootstrap as discussed earlier. 
 
 There are multiple lines of inquiry that lead out from this point. We wind up by listing some of these:   
 \begin{enumerate}
 
 \item
 Since we have the collinear $\langle J_0 J_0 J_0 J_0 \rangle$, $\langle J_1 J_1 J_0 J_0 \rangle $, and $\langle J_1 J_1 J_1 J_1 \rangle $ in the CS + free fermionic theory, we may immediately  compute the corresponding correlators in CS + critical fermionic theory via a Legendre transform (as was done in \cite{Bedhotiya:2015uga} and \cite{Turiaci:2018dht}). Assuming the bosonisation dualities, we then have these correlators in all the quasi-fermionic and -bosonic theories. Note that $\langle J_1 J_1 J_1 J_1 \rangle $ does not change under this transformation.

 \item 
 % CS Bosonic theories
 We may also use the brute force approach discussed here to directly compute the correlators in the CS + bosonic theories. Comparing these with our results would then give more evidence toward the bosonisation dualities. However, the computation there is slightly more complicated due to the fact that the spin-1 current contains gauge bosons and hence the contributing diagrams are different.
% The brute force approach discussed here may be replicated for the quasi-bosonic theories. However, the situation there is slightly more complicated due to the fact that the spin-1 current contains gauge bosons and hence the contributing diagrams are different. What we can do at the moment, is to compute $\langle J_0 J_0 J_0 J_0 \rangle $. This was also done in \cite{Turiaci:2018dht}, and as in the fermionic case, they found the result with the aid of the inversion formula (and numerics). We may be able to obtain an analytic expression (again, in the collinear limit).
   
 \item 
 %Higher correlators a la Yacoby
 
 As was mentioned in the introduction, \cite{Yacoby:2018yvy} found a way to compute all scalar correlators in the bosonic theory, in the collinear limit. It may be possible to replicate this approach in the fermionic case and hence also compute higher-point correlators involving $J_0$, and perhaps also $J_1$.
 
 \item

 This approach (and also perhaps the one of \cite{Yacoby:2018yvy}) should also be tried on correlators involving higher spin currents. As was mentioned earlier, the main problem is that for $s > 1$ in fermionic theories, and $s > 0$ in bosonic theories, $J_s$ contains gauge bosons, and hence the necessary ingredients haven't been worked out yet.

 \item 
 % Is info from collinear limit enough to fix the rest? (as in 

Our computations are in the collinear limit and, on their own, don't fix the correlator; other tools like the inversion formula must be employed to bootstrap the full correlator. Unlike in \cite{Turiaci:2018dht} and \cite{Li:2019twz}, our result contains extra structures that do not appear in the limiting ($ \tilde {\lambda } \rightarrow 0, \infty $) free fermion, and critical scalar theories; and are not yet known for general values of momenta. They can, however, be computed perturbatively.

As discussed in the introduction, if an analysis akin to that of \cite{Turiaci:2018dht} is employed in other cases, the collinear result could be used to the fix resultant ambiguities and hence nail down the full result. For this approach, it is necessary that the contact diagram corrections not vanish in the collinear limit. Our computations show that this is indeed the case for $ \langle J_1 J_1 J_1 J_1 \rangle $, and probably also for $ \langle J_0 J_0 J_1 J_1 \rangle $. 

% For instance, in the case of $ \langle J_1 J_1 J_1 J_1 \rangle $, the possible corrections are contact Witten diagrams with vertices $ \left( F^2  \right)^2 $,  $ F^4 $, and $ F^{\mu \nu } \left( \partial _{\mu}  F_{\rho} ^{\sigma} \right) \left( \partial _{\nu} F_{\sigma} ^{\lambda } \right) F_{\lambda} ^{\rho}$. We have found (see Appendix \ref{apd:contact}) that the contribution from the latter vertex vanishes in the collinear limit, and hence something more is needed to get at the full result.

 \item 
 %Analyticity in spin: does the naive continuation give the same result?
 
It is also interesting to study the possibility of other correlators in our theories being analytic in spin (as was speculated on by \cite{Turiaci:2018dht}). It is not yet clear precisely what this property means, so CS matter theories could potentially become a playground for understanding the inversion formula better.

\item 
%Higher-spin Ward identities  
  
  The work of \cite{Li:2019twz} has opened up another avenue to perform computations in these theories. Their approach uses nothing but the anomalous higher-spin Ward identities and works them into a perturbation-series-like form, but in $\tilde{ \lambda} $. Since the arguments we gave earlier in the section lead us to expect that all four-point functions of single-trace operators (normalised suitably) have a finite expansion in this $\tilde{ \lambda} $, this line of attack could be especially effective. This could also provide more conclusive evidence to answer the question of whether weakly broken higher-spin symmetry could uniquely fix all the correlators in CS-matter theories. One drawback of this method is that it may be insensitive to contact terms \footnote{This time, we mean delta functions in position space.}, since correlators are written in position space.
%  
%    Although our theories are constrained by higher-spin symmetry, it is not yet clear how to exploit this fact. These identities are algebraic in momentum space and using these could let us compute higher-spin (involving operators of higher spin) correlation functions from lower ones. As was discussed earlier, there is also the question of whether these (along with the inversion formula) uniquely fix all the correlators in these theories.  
  
 \item 
 %Vasiliev four-point functions
 
 A long term goal would be to compute all correlators in CS matter theories and match them with those in Vasiliev theories. But computations in the bulk are very hard to do and so far, only some three-point functions have been computed \cite{Giombi:2012ms} and compared with those in the dual theory.

 \end{enumerate}

\section{Acknowledgements}
We thank Ofer Aharony for coming up with the idea for this project, many helpful discussions, and a thorough reading of the manuscript. We thank Ran Yacoby and Shai Chester for useful discussions. We thank Efi Efrati for help with the computation. This work was supported in part by an Israel Science Foundation center for excellence grant (grant number 1989/14) and by the Minerva foundation with funding from the Federal German Ministry for Education and Research.

 \appendix

% Appendix  1--------------------------------------------------------------------------

\section{The four-fermion vertex \label{apd:four_vertex}}

The four-fermion vertex (Fig. \ref{fig:ladint}) was first computed by \cite{Inbasekar:2015tsa}, and makes an appearance in the `ladder' diagrams (see Fig. \ref{fig:efflad}) that contribute to the four-point function. We give a rough sketch of our derivation, following \cite{Inbasekar:2015tsa} and \cite{Turiaci:2018dht}, but using the conventions of \cite{GurAri:2012is}. This means that our expressions are different by a few minus signs.

The sum of the ladder-type diagrams (with insertions $\mathcal{O}_i$, that stand in for $V(q,p)$ and $V^{\pm}(q,p)$) may be arranged into the series shown in Fig. \ref{fig:efflad} (where the vertices and fermion propagators are the all-orders versions, and the zipline is the gluon propagator). This series then corresponds to the integral:
\begin{multline}
- \int \frac{\text{d}^3 r}{(2\pi )^3} \frac{\text{d}^3 k}{(2\pi )^3}   \text{Tr} \left[ \text{S}(k + q_1 + q_2) \mathcal{O}_2 \text{S}(k + q_1) \mathcal{O}_1 \text{S}(k) \gamma ^{\mu} \text{S}(r) \mathcal{O}_4 \text{S}(r - q_4) \mathcal{O}_3 \text{S}(r - q_3 - q_4) \gamma ^{\nu} \right] \text{G}_{\mu \nu}(k-r) \\
+ \nonumber \\
- \int 
 \frac{\text{d}^3 r}{(2\pi )^3} \frac{\text{d}^3 k}{(2\pi )^3}    \frac{\text{d}^3 l}{(2\pi )^3} \text{Tr} \left[ \text{S}(k + q_1 + q_2) \mathcal{O}_2 \text{S}(k + q_1) \mathcal{O}_1 \text{S}(k) \gamma ^{\mu} \text{S}(l) \gamma ^{\rho} \text{S}(r) \mathcal{O}_4 \text{S}(r - q_4) \mathcal{O}_3 \right. \\ 
 \left. \text{S}(r - q_3 - q_4) \gamma ^{\sigma} \text{S}(l + q_1 + q_2) \gamma ^{\nu} \right] \text{G}_{\mu \nu}(k-l) \text{G}_{\rho \sigma}(l-r) \nonumber \\
+ ... \\ \nonumber
\end{multline}
where we have omitted the momentum dependences of the vertices. Restoring the fermion indices lets us factor out the ladder part.

\begin{multline}
- \int \frac{\text{d}^3 r}{(2\pi )^3} \frac{\text{d}^3 k}{(2\pi )^3}    \left[ \text{S}(k + q_1 + q_2) \mathcal{O}_2 \text{S}(k + q_1) \mathcal{O}_1 \text{S}(k)\right] _j ^m \left( \gamma ^{\mu} \right)_m ^n \left[ \text{S}(r) \mathcal{O}_4 \text{S}(r - q_4) \mathcal{O}_3 \text{S}(r - q_3 - q_4) \right] _n ^i  \nonumber \\
\left( \gamma ^{\nu} \right) _i ^j  \text{G}_{\mu \nu}(k-r) \nonumber \\
+ \\
- \int 
 \frac{\text{d}^3 r}{(2\pi )^3} \frac{\text{d}^3 k}{(2\pi )^3}    \frac{\text{d}^3 l}{(2\pi )^3} \left[ \text{S}(k + q_1 + q_2) \mathcal{O}_2 \text{S}(k + q_1) \mathcal{O}_1 \text{S}(k)\right] _j ^m \left[ \gamma ^{\mu} \text{S}(l) \gamma ^{\rho} \right]_m ^n \left[ \text{S}(r) \mathcal{O}_4 \text{S}(r - q_4) \mathcal{O}_3 \right.  \\
\left. \text{S}(r - q_3 - q_4) \right] _n ^i \left[ \gamma ^{\sigma} \text{S}(l + q_1 + q_2) \gamma ^{\nu} \right] _i ^j
 \ \text{G}_{\mu \nu}(k-l) \text{G}_{\rho \sigma}(l-r)  \nonumber \\
+ ... \\ \nonumber
\end{multline}
\begin{multline}
\begin{gathered}
= - \int \frac{\text{d}^3 r}{(2\pi )^3} \frac{\text{d}^3 k}{(2\pi )^3}     \left[ \text{S}(k + q_1 + q_2) \mathcal{O}_2 \text{S}(k + q_1) \mathcal{O}_1 \text{S}(k)\right] _j ^m \\
\left\{ \left( \gamma ^{\mu} \right)_m ^n  \left( \gamma ^{\nu} \right) _i ^j \text{G}_{\mu \nu}(k-r) + \int \frac{\text{d}^3 l}{(2\pi )^3} \left[ \gamma ^{\mu} \text{S}(l) \gamma ^{\rho} \right]_m ^n \left[ \gamma ^{\sigma} \text{S}(l + q_1 + q_2) \gamma ^{\nu} \right] _i ^j  \text{G}_{\mu \nu}(k-l) \text{G}_{\rho \sigma}(l-r) + ... \right\} \\
\left[ \text{S}(r) \mathcal{O}_4 \text{S}(r - q_4) \mathcal{O}_3 \text{S}(r - q_3 - q_4) \right] _n ^i  
\end{gathered}
\end{multline}
\begin{gather}
= - \int \frac{\text{d}^3 r}{(2\pi )^3} \frac{\text{d}^3 k}{(2\pi )^3}     \left[ \text{S}(k + q_1 + q_2) \mathcal{O}_2 \text{S}(k + q_1) \mathcal{O}_1 \text{S}(k)\right] _j ^m \Gamma _{im} ^{jn} (k,q_1 + q_2,r)
\left[ \text{S}(r) \mathcal{O}_4 \text{S}(r - q_4) \right. \nonumber \\
\left. \mathcal{O}_3 \text{S}(r - q_3 - q_4) \right] _n ^i .
\end{gather}
In the last line, we have defined $\Gamma _{im} ^{jn} (k,q,r)$, the four-fermion effective action or the ladder. Assuming that $q$ lies along the z-axis, the resummation may be cast into a recursive Schwinger-Dyson equation, and solved following the methods of \cite{Inbasekar:2015tsa}.
\begin{figure}
    \centering
\begin{IEEEeqnarray}{rCl}
 \raisebox{-12ex}{\begin{tikzpicture}
\begin{feynman}
\vertex (a1) ;
\vertex[right=2cm of a1] (a2);
\vertex[above=4cm of a1] (b1);
\vertex[right=2cm of b1] (b2);
\vertex[above=2cm of a1] (ctemp1);
\vertex[above=2cm of a2] (ctemp2);
\vertex[right=1cm of ctemp1] (ctemp);
\fill[black] (ctemp)  circle (1.5 cm and 0.5 cm);
\diagram*[small]{
(a1) -- [fermion, , momentum=\(r\) ] (ctemp1),
(ctemp1) -- [momentum=\(k\)] (b1),
(b2) -- [momentum=\(k + q\)] (ctemp2),
(ctemp2) --  [fermion, momentum=\(r + q\)] (a2),
};
\end{feynman}
\end{tikzpicture}} \quad \quad
& = &
\quad \quad
 \raisebox{-12ex}{\begin{tikzpicture}
\begin{feynman}
\vertex (a1) ;
\vertex[right=2cm of a1] (a2);
\vertex[above=4cm of a1] (b1);
\vertex[right=2cm of b1] (b2);
\vertex[above=2cm of a1] (ctemp1);
\vertex[above=2cm of a2] (ctemp2);
\vertex[right=1cm of ctemp1] (ctemp);
\diagram*[small] {
(a1) -- [fermion] (ctemp1),
(ctemp1) -- (b1),
(b2) --(ctemp2),
(ctemp2) --  [fermion] (a2),
(ctemp1) -- [photon] (ctemp2)
};
\end{feynman}
\end{tikzpicture}} \quad \quad
+
\quad \quad
 \raisebox{-12ex}{\begin{tikzpicture}
\begin{feynman}
\vertex (a1) ;
\vertex[right=2cm of a1] (a2);
\vertex[above=4cm of a1] (b1);
\vertex[right=2cm of b1] (b2);
\vertex[above=1.33cm of a1] (ctemp1);
\vertex[above=1.33cm of a2] (ctemp2);
\vertex[above=1.33cm of ctemp1] (ctemp11);
\vertex[above=1.33cm of ctemp2] (ctemp21);
\vertex[above=1.66cm of a1, blob] (ctemp3) {};
\vertex[above=1.66cm of a2, blob] (ctemp3) {};
\diagram*[small] {
(a1) -- [fermion] (ctemp1),
(ctemp1) --  (ctemp11),
(ctemp11) -- (b1),
(b2) --  (ctemp21),
(ctemp21) --  (ctemp2),
(ctemp2) --  [fermion] (a2),
(ctemp1) -- [photon] (ctemp2),
(ctemp11) -- [photon] (ctemp21)
};
\end{feynman}
\end{tikzpicture}} \quad \quad
+
\quad \quad ... 
\end{IEEEeqnarray}
    \caption{Effective four-fermion interaction.}
    \label{fig:ladint}
\end{figure}
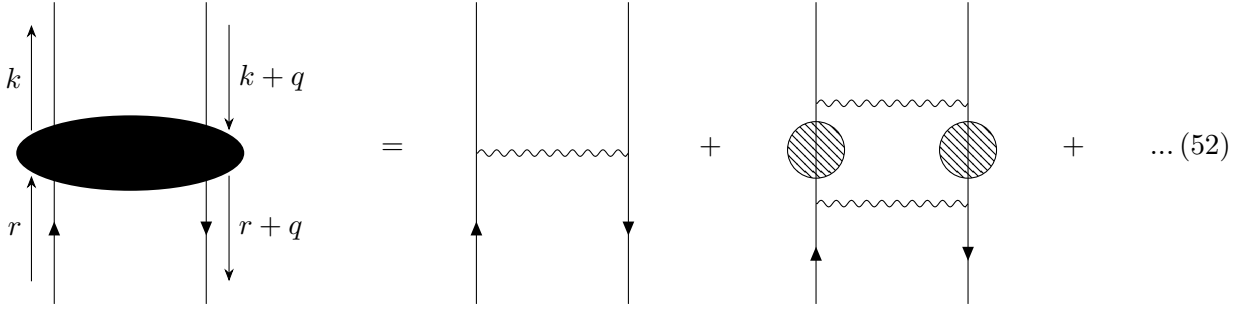
\begin{center}
\begin{figure}
    \centering
    \includegraphics[width=0.80\textwidth]{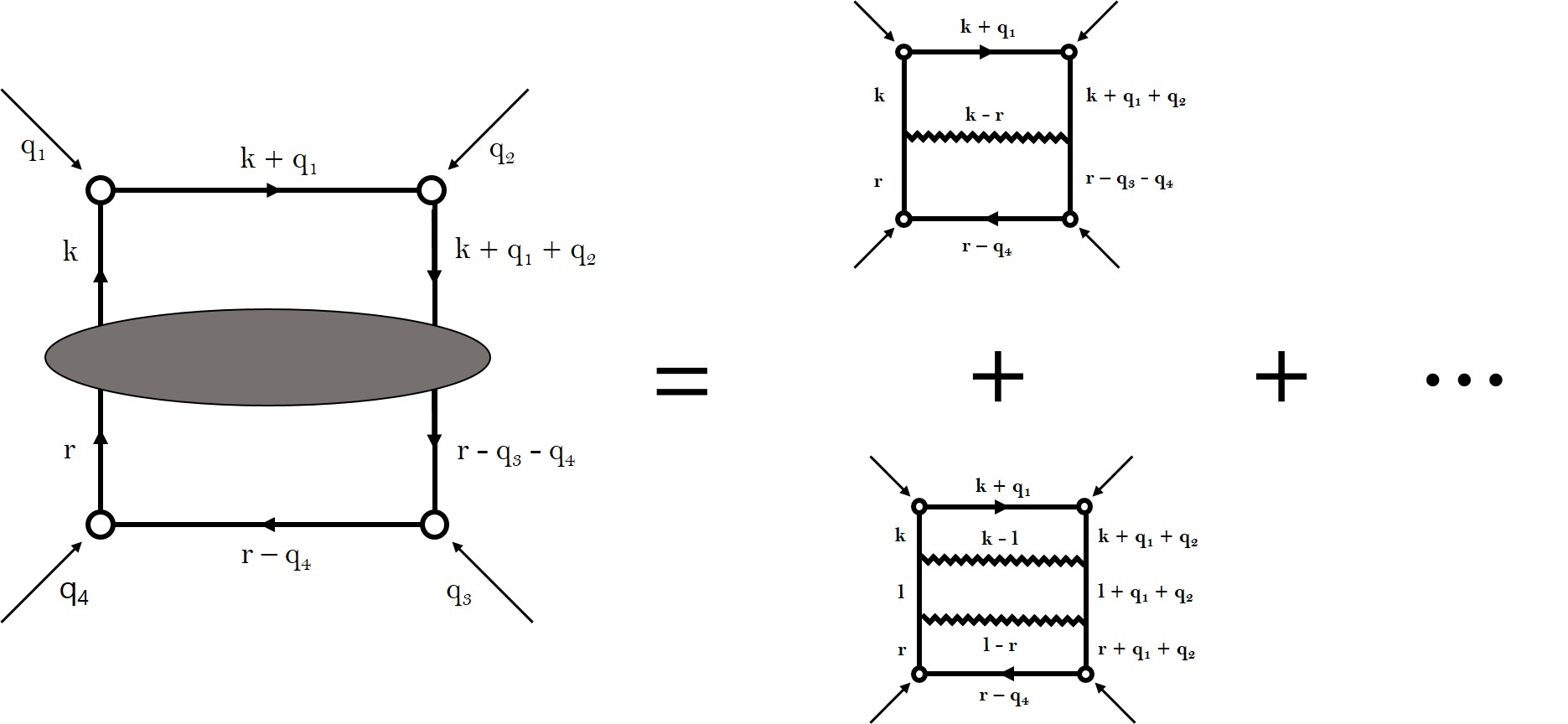},
    \caption{The sum of ladder-type diagrams. The fermion lines stand for exact fermionic propagators, and the ziplines stand for gluon propagators.}
\label{fig:efflad}
\end{figure}
\end{center}
Once we have $\Gamma _{im} ^{jn} (k,q,r)$, we recast it in a different basis following \cite{Turiaci:2018dht}. Since $\Gamma _{im} ^{jn} (k,q,r)$ is a $2 \times 2 \times 2 \times 2 $ matrix, it may be expressed as:
\begin{equation}
\Gamma _{im} ^{jn} (k,q,r) = \Gamma _{AB} (k,q,r) \left( \gamma ^A \right)_m ^n \left( \gamma ^B \right)_i ^j,
\end{equation}
where $A,\ B \in \{+, -, 3, I \} $, and  $\gamma ^I = \mathbf{1}_{2 \times 2}$.\footnote{The definition may be inverted to give
\[
\Gamma _{AB} (k,q,r) = \frac{1}{4} \left( \gamma _{A} \right) _n ^m \left( \gamma _{B} \right) _j ^i \Gamma _{im} ^{jn} (k,q,r),
\]
which is how $\Gamma _{AB} (k,q,r)$ is computed from the effective action.}
Using this, the ladder-type diagrams may be rexpressed as:
\begin{equation}
\begin{gathered}
- \int \frac{\text{d}^3 r}{(2\pi )^3} \frac{\text{d}^3 k}{(2\pi )^3}     \left[ \text{S}(k + q_1 + q_2) \mathcal{O}_2 \text{S}(k + q_1) \mathcal{O}_1 \text{S}(k)\right] _j ^m \Gamma _{AB} (k,q_1 + q_2,r) \left( \gamma ^A \right)_m ^n \left( \gamma ^B \right)_i ^j \\
\left[ \text{S}(r) \mathcal{O}_4 \text{S}(r - q_4) \mathcal{O}_3 \text{S}(r - q_3 - q_4) \right] _n ^i \\ 
= - \int \frac{\text{d}^3 r}{(2\pi )^3} \frac{\text{d}^3 k}{(2\pi )^3}     \text{Tr} \left[ \text{S}(k + q_1 + q_2) \mathcal{O}_2 \text{S}(k + q_1) \mathcal{O}_1 \text{S}(k)\gamma ^A \text{S}(r) \mathcal{O}_4 \text{S}(r - q_4) \mathcal{O}_3 \text{S}(r - q_3 - q_4) \gamma ^B \right]  \\
\Gamma _{AB} (k,q_1 + q_2,r) \\ 
\end{gathered}
\end{equation}
making it easy to see the Feynman rules for the ladder vertex. In this basis, (the non-vanishing components of) the ladder is given by:

\begin{IEEEeqnarray}{rCl}
\Gamma _{+ +}(k, q, r)  & = &  -\left( \frac{ r^- }{r_s ^2 \left( r^+ - k^+  \right) } B_0 + \frac{ k^- }{2 \left( r^+ - k^+  \right) } B_1 \right)  \\
\Gamma _{+ I}(k, q, r)  & = & \frac{ k^- r^+ }{2 \left( r^+ - k^+  \right) } B_2 + \frac{ 1 }{2 \left( r^+ - k^+  \right) } B_3  \\
\Gamma _{I +}(k, q, r)  & = & -\left( \frac{1 }{2 \left( r^+ - k^+  \right) } W_3 - \frac{ r^- }{r_s ^2 } W_2 \right)  \\
\Gamma _{I I}(k, q, r)  & = & - \frac{ r^+ }{2 \left( r^+ - k^+  \right) } W_0 + \frac{ 1 }{2 } W_1 
\end{IEEEeqnarray} 
where the $W_i$s and $B_i$s are functions of dimensionless combinations of momenta $y = \frac{2 k_s}{\left| q \right|}$, $z = \frac{2 r_s}{\left| q \right|}$, $\Lambda ' = \frac{2 \Lambda}{\left| q \right|}$, and of the ``parity-invariant" coupling $\widehat{\lambda} = \text{sgn}(q)\lambda $  :
\begin{IEEEeqnarray}{rCl}
W_0  & = & 4 \iu \pi \widehat{\lambda} \frac{ \left( -1 + e^{-2 \iu \widehat{\lambda} \left( \text{tan} ^{-1} (z) - \text{tan}^{-1} (y) \right) } \right) }{\left| q \right|}    \\
W_1  & = & 4 \iu \pi \widehat{\lambda} \frac{ \left( -1 + e^{-2 \iu \widehat{\lambda} \left( \text{tan} ^{-1} (\Lambda ') - \text{tan}^{-1} (y) \right) } \right) \left( 1 + e^{- 2 \iu \widehat{\lambda}\  \text{tan} ^{-1} (z)} \right) }{\left| q \right| \left( 1 + e^{- 2 \iu \widehat{\lambda}\  \text{tan} ^{-1} (\Lambda ')} \right)}  \\
W_2  & = & -2 \pi \widehat{\lambda}\ \text{sgn}(q) \frac{ \left( -1 + e^{- 2 \iu \widehat{\lambda} \left( \text{tan} ^{-1} (\Lambda ') - \text{tan}^{-1} (y) \right) } \right) }{\left( 1 + e^{- 2 \iu \widehat{\lambda}\  \text{tan} ^{-1} (\Lambda ')}  \right)} \left( - \widehat{\lambda} z - \iu + \left( - \widehat{\lambda} z + \iu \right) e^{- 2 \iu \widehat{\lambda}\ \text{tan}^{-1} (z) } \right)  \\
W_3  & = & 2 \pi \widehat{\lambda}\ \text{sgn}(q) \left( -\left( - \widehat{\lambda} z + \iu \right) e^{- 2 \iu \widehat{\lambda}\left( \text{tan} ^{-1} (z) - \text{tan}^{-1} (y) \right)} -  \widehat{\lambda} z - \iu \right)  \\
& & \nonumber  \\
B_0 & = & - \frac{\iu  \pi  \widehat{\lambda} \left| q \right| }{ 1 + e^{- 2 \iu \widehat{\lambda}\ \text{tan} ^{-1} (\Lambda ')}} e^{ 2 \iu \widehat{\lambda}\ \text{tan} ^{-1} (y)} \left( -1 + \iu z \widehat{\lambda} + e^{- 2 \iu \widehat{\lambda}\ \text{tan} ^{-1} (z)} \left( 1 + \iu z \widehat{\lambda} \right)  \right)  \nonumber   \\
& & \ \ \ \ \ \ \ \ \ \ \ \ \ \ \left(  e^{- 2 \iu \widehat{\lambda}\ \text{tan} ^{-1} (y)} \left( 1 + \iu y \widehat{\lambda} \right) + e^{-2 \iu \widehat{\lambda}\ \text{tan} ^{-1} (\Lambda ')} \left( 1 - \iu y \widehat{\lambda} \right)  \right)  \\
B_1 & = &  - \frac{8 \iu  \pi  \widehat{\lambda} }{\left( 1 + e^{-2 \iu \widehat{\lambda}\ \text{tan} ^{-1} (\Lambda ')} \right) y^2 \left| q \right|} e^{ 2 \iu \widehat{\lambda}\ \text{tan} ^{-1} (y)} \left( e^{- 2 \iu \widehat{\lambda}\ \text{tan} ^{-1} (z)} \left( 1 + \iu z \widehat{\lambda} \right) + e^{- 2 \iu \widehat{\lambda}\ \text{tan} ^{-1} (\Lambda ')} \left( 1 - \iu z \widehat{\lambda} \right)  \right) \nonumber  \\
& & \ \ \ \ \ \ \ \ \ \ \ \ \ \  \left( 1 - \iu y \widehat{\lambda} - e^{-2 \iu \widehat{\lambda}\ \text{tan} ^{-1} (y)} \left( 1 + \iu y \widehat{\lambda} \right)  \right)  \\
B_2 & = & -\frac{16 \iu  \pi  \widehat{\lambda}\ \text{sgn}(q) }{\left( 1 + e^{- 2 \iu \widehat{\lambda}\ \text{tan} ^{-1} (\Lambda ')} \right) y^2 \left| q \right|^2 }  e^{ 2 \iu \widehat{\lambda}\ \text{tan} ^{-1} (y)} \left(  e^{- 2 \iu \widehat{\lambda}\ \text{tan} ^{-1} (\Lambda ')} -  e^{- 2 \iu \widehat{\lambda}\ \text{tan} ^{-1} (z)} \right)\nonumber \\
& &   \left( -1 + \iu y \widehat{\lambda} + e^{- 2 \iu \widehat{\lambda}\ \text{tan} ^{-1} (y)} \left( 1 + \iu y \widehat{\lambda} \right)  \right)   \\
B_3 & = & - \frac{2 \iu  \pi  \widehat{\lambda}\ \text{sgn}(q)}{ 1 + e^{- 2 \iu \widehat{\lambda}\ \text{tan} ^{-1} (\Lambda ')}} e^{ 2 \iu \widehat{\lambda}\ \text{tan} ^{-1} (y)} \left( 1 + e^{ - 2 \iu \widehat{\lambda}\ \text{tan} ^{-1} (z)} \right) \nonumber \\
& & \left(  e^{- 2 \iu \widehat{\lambda}\ \text{tan} ^{-1} (y)} \left( 1 + \iu y \widehat{\lambda} \right) + e^{- 2 \iu \widehat{\lambda}\ \text{tan} ^{-1} (\Lambda ')} \left( 1 - \iu y \widehat{\lambda} \right)  \right) 
\end{IEEEeqnarray}

% Appendix 2--------------------------------------------------------------------------

\section{ $\left< J_1 J_1 J_1 J_1 \right> $: the full expression \label{apd:j1j1j1j1}}

We give here the full expression for the $\langle J_1 J_1 J_1 J_1 \rangle $ correlator:
\begin{IEEEeqnarray}{C}
\langle \langle \tilde{J}_1^+ \newline (q + \epsilon + \delta ) \tilde{J}_1^- (-q ) \tilde{J}_1^+ ( -\epsilon ) \tilde{J}_1^- (- \delta ) \rangle \rangle = \frac{1}{\tilde{N}} \left( \frac{1}{(1 + \tilde{\lambda}^2)^2} F(q,\epsilon, \delta) 
+ \frac{\tilde{\lambda}}{(1 + \tilde{\lambda}^2)^2} O(q,\epsilon, \delta) \right.
\nonumber \\
 + \frac{\tilde{\lambda}^2}{(1 + \tilde{\lambda}^2)^2} ( F(q,\epsilon, \delta) + B(q,\epsilon, \delta) )
  + \frac{\tilde{\lambda}^3}{(1 + \tilde{\lambda}^2)^2} O(q,\epsilon, \delta)
   + \left. \frac{\tilde{\lambda}^4}{(1 + \tilde{\lambda}^2)^2} B(q,\epsilon, \delta)  \right) \\
   = \frac{1}{\tilde{N}} \left(  \frac{1}{1 + \tilde{\lambda}^2} F(q,\epsilon, \delta) 
+ \frac{\tilde{\lambda}}{1 + \tilde{\lambda}^2} O(q,\epsilon, \delta)
 + \frac{\tilde{\lambda}^2}{1 + \tilde{\lambda}^2} B(q,\epsilon, \delta) \right)
\end{IEEEeqnarray}
where:
\begin{IEEEeqnarray}{rCl}
8 F(q, \epsilon, \delta ) & = & 
 \text{sgn} \left( q \right) \frac{ q\left( 2 \de ^2 \e \left( \de + \e \right) + q ^3 \left (\de + \e \right )+ q ^2 \left( 3 \de ^2 + 5 \de \e + \e ^2 \right) + 2 \de  q\left( \de + \e \right) \left( \de + 2 \e \right) \right) }{ \de \e \left( \de + \e \right) \left( \de + q \right) \left( q + \e \right) \qp } \nonumber \\
 & & + 
 \text{sgn} \left( \e \right) \frac{ \e \left( - \e^2 \left( 3 \de^2 + 3 q^2 + 5 \de q \right) - \e ^3 \left( \de + q \right) - 2 \e \left( \de + q \right)^3 - 2 \de q \left( \de + q \right)^2 \right)  }{ \de q \left( \de + \e \right) \left( \de + q \right) \left( q  +\e \right) \qp } \nonumber \\
& & +
 \text{sgn} \left( \de \right) 
 \frac{ \de \left( \de ^2 \e \left( \de + \e \right) + 2 q^3 \left( \de + \e \right) + q^2 \left( 3 \de ^2 + 6 \de \e + 2 \e^2 \right) + \de q \left( \de + \e \right) \left( \de + 4 \e \right) \right)}{ \e q \left( \de + \e \right) \left( \de + q \right) \left( q  +\e \right) \qp } \nonumber \\
 & & +
 \text{sgn} \qp \frac{ \qp \left( \de \e \left( \de - \e \right) \left( \de + \e \right) + q^3 \left( \de + \e \right) + \de q^2 \left( 2 \de + \e \right) + q \left( \de - \e \right) \left( \de + \e \right)^2 \right)  }{ \de \e q \left( \de + \e \right) \left( \de + q \right) \left( q  +\e \right)  } \nonumber \\
 &  & -
   \text{sgn} \left( q + \e \right) \frac{ \left( q - \e \right) \left( q + \e \right) \left( 2 \de + q + \e \right)}{ \de q \e \qp}
   -
   \text{sgn} \left( q + \de \right) \frac{ \left( \de + q \right)^3 }{ \de q \e \qp}
   \nonumber \\
   &  & -
   \text{sgn} \left( \e + \de \right) \frac{ \left( \de + \e \right) \left( \de - \e \right) \left( 2q + \de + \e \right) }{ \de q \e \qp} 
\end{IEEEeqnarray}

\begin{IEEEeqnarray}{rCl}
-8 \iu O(q, \epsilon , \delta ) & = &
\frac{ \e ^2 \left( \de ^2 + q^2 \right) + \e \left( \de ^2 + q^2 \right) \left( \de + q \right) + \left( \de ^2 + q ^2 + \de q \right) \left( \de + q \right)^2  }{ \de q \e \left( \de + q \right) \qp } 
-
\text{sgn} \left( q \right) \text{sgn} \left( \e + q \right)
\frac{ q \left( 2 \de + q + \e \right)}{ \de \e \qp }
\nonumber \\
& &  -
\text{sgn} \left( \e \right) \text{sgn} \left( \e + q \right) 
\frac{ \e \left( 2 \de + q + \e \right)}{ \de q \qp }
+
\text{sgn} \left( \de \right) \text{sgn} \left( \e + q \right) 
\frac{ \de \left( q - \e \right)}{ \e q \qp }
\nonumber \\
& & -
\text{sgn} \left( q + \de + \e \right) \text{sgn} \left( \e + q \right) 
\frac{ \left( q - \e \right) \left( q + \de + \e \right)}{ \e q \de } 
 +
\text{sgn} \left( \de + \e \right) \text{sgn} \left(  q \right)
\frac{ q \left( \de - \e \right)}{  \de \e \left( q + \de + \e \right)}
\nonumber \\
& & -
\text{sgn} \left( \de + \e \right) \text{sgn} \left(  \e \right)
\frac{ \e \left( \de +2q + \e \right)}{  \de q \left( q + \de + \e \right)} 
-
\text{sgn} \left( \de + \e \right) \text{sgn} \left(  \de \right)
\frac{ \de \left( \de + 2q +  \e \right)}{  q \e \left( q + \de + \e \right)}
\nonumber \\
& & -
\text{sgn} \left( \de + \e \right) \text{sgn} \left(  q + \e + \de \right)
\frac{ \left( \de - \e \right) \qp}{  \de \e  q } 
 +
\text{sgn} \left( q \right) \text{sgn} \left(  q + \e + \de \right)
\frac{ q \qp}{  \de \e  \left( q + \e \right) }
\nonumber \\
& & -
 \text{sgn} \left( \e \right) \text{sgn} \left(  q + \e + \de \right)
 \frac{ \e \left( \de ^2 + q^2 \right) \qp }{  \de q \left( \de + \e \right) \left( \de + q \right) \left( q + \e \right)}
 +
 \text{sgn} \left( \de \right) \text{sgn} \left(  q + \e + \de \right)
 \frac{ \de \qp }{ q \e \left( \de + \e \right) }
 \nonumber \\
& & -
 \text{sgn} \left( q \right) \text{sgn} \left( \de \right)
 \frac{ \de q \left( 2 \e \left (\de + q \right) + \left( \de  + q \right)^2 + 2 \e ^2 \right)}{  \e \left( \de + \e \right) \left( \de + q \right) \left( q + \e \right) \qp }
  +
 \text{sgn} \left( \e \right) \text{sgn} \left( \de \right)
 \frac{ \de \e }{  q \left( q + \e \right) \qp }
\nonumber \\
& &  +
 \text{sgn} \left( \e \right) \text{sgn} \left( q \right)
 \frac{ q \e }{  \de \left( \de + \e \right) \qp }
\end{IEEEeqnarray}

\begin{IEEEeqnarray}{rCl}
8 B(q, \epsilon, \delta ) & = &
 \text{sgn} \left( q  \right) 
\frac{ \de q \left( \left( q +\e \right) \left( \de + \e \right) + q^2 \right)  }{ \e \left( \de + \e \right) \left( \de + q \right) \left( q + \e \right) \qp } 
-
\text{sgn} \left( \e  \right)
\frac{ \e \qp \left( \de \e + q \left( \de + \e \right) \right) }{ \de q \left( \de + \e \right) \left( \de + q \right) \left( q + \e \right) }
\nonumber \\
& & 
+
\text{sgn} \left( \de  \right)
\frac{ \de q \left( \left( q + \de \right) \left( \de + \e \right) + \e ^2 \right) }{ \e \left( \de + \e \right) \left( \de + q \right) \left( q + \e \right) \qp }
-
\text{sgn} \qp
\frac{ \e \qp \left( \left( \de + \e \right) \left( q + \de \right)  + q^2 \right)}{ \de q \left( \de + \e \right) \left( \de + q \right) \left( q + \e \right) }
\nonumber \\
& &
+
\text{sgn} \left( q  \right) \text{sgn} \left( q + \e  \right)
\text{sgn} \qp
\frac{ q \qp}{ \de \e \left( q + \e \right) }
-
\text{sgn} \left( q  \right)
\text{sgn} \left( q + \e  \right)
\text{sgn} \left( \de  \right)
\frac{ \de q }{ \e \left( q + \e \right) \qp }
\nonumber \\
& &
-
\text{sgn} \left( q  \right)
\text{sgn} \left( \de + \e  \right)
\text{sgn} \left( \de  \right)
\frac{ \de q }{ \e \left( \de + \e \right) \qp }
-
\text{sgn} \left( q  \right)
\text{sgn} \left( \de + \e  \right)
\text{sgn} \left( \e  \right)
\frac{ \e q }{ \de \left( \de + \e \right) \qp }
\nonumber \\\
&  &
+
\text{sgn} \left( \e  \right)
\text{sgn} \left( q + \e  \right)
\text{sgn} \qp
\frac{ \e \qp }{ \de  q \left( q + \e \right) }
+
\text{sgn} \left( \de  \right)
\text{sgn} \left( \de + \e  \right)
\text{sgn} \qp
\frac{ \de \qp }{ \e  q \left( \de + \e \right) }
\nonumber \\
& &
+
\text{sgn} \left( \e  \right)
\text{sgn} \left( \de + \e  \right)
\text{sgn} \qp
\frac{ \e \qp }{ \de  q \left( \de + \e \right) }
-
\text{sgn} \left( \de  \right)
\text{sgn} \left( \e  \right)
\text{sgn} \left( q + \e \right)
\frac{ \de \e }{ q  \left( q + \e \right) \left( q + \de + \e \right) }
\nonumber \\
& &
-
\text{sgn} \left( q + \de  \right)
\frac{ \left( \de + q \right)^3 }{ q \de \e \qp }
\end{IEEEeqnarray}
are respectively, the free fermionic, odd, and critical bosonic structures.

\bibliographystyle{jhep}    
\bibliography{library}

\end{document}